\documentclass[pre,reprint,aps,superscriptaddress,showpacs,floatfix]{revtex4-2}
\usepackage[dvipdfmx]{graphicx}
\usepackage{color}
\usepackage{bm,amsmath,amssymb,multirow,CJK,grffile}
\usepackage{mathrsfs}
\usepackage[title]{appendix}
\usepackage{braket}
\usepackage{array}
\makeatletter
\let\MYcaption\@makecaption
\makeatother

\usepackage{subcaption}
\makeatletter
\let\@makecaption\MYcaption
\makeatother
\usepackage[inline]{enumitem}
 
\begin{document}
\begin{CJK*}{UTF8}{}
	\title{
	Defect configuration of an active nematic around a circular obstacle
}
 	\author{Hiroki Matsukiyo
}
\affiliation{
	Department of Physics, Kyushu University, 744 Motooka, Nishi-ku, Fukuoka 819-0395, Japan
}
\author{Jun-ichi Fukuda
}
\affiliation{
	Department of Physics, Kyushu University, 744 Motooka, Nishi-ku, Fukuoka 819-0395, Japan
}
  	\date{\today}
	\begin{abstract}
To enhance the understanding of the behavior of active nematic,
	it is important to understand the behavior of topological defects.
	In this paper, we study the configuration of topological defects
	of a two-dimensional active nematic around a circular obstacle.
	In the case of a passive
	nematic liquid crystal, the equilibrium configuration of defects can be easily
	identified by
	the method of image charges.
	In the case of an active nematic, however, one must take account of the flow field
	generated by active constituents,
	and the problem of identifying the defect configuration becomes complicated.
We first
	perform numerical simulations
	and investigate how the stationary defect configuration deviates from the passive case.
	Furthermore, we carry out a theoretical calculation
	based on an analytical expression relating the defect velocity with the force exerted on the defect.
	Our theoretical calculation qualitatively reproduces the simulation results.
	Our study may be applied to describing the behaviour of e.g. cell populations in the presence of obstacles,
	and has the potential to benefit related fields, e.g., developmental biology.
\end{abstract}
 	\maketitle
\end{CJK*}
\section{Introduction}
\label{introduction}
Active matter refers to systems composed of (usually a large number of)
self-propelled agents, e.g.,
flock of birds, school of fish, bacterial colony,
etc \cite{marchetti2013hydrodynamics}.
Constituents of active matter consume chemical energy, move while interacting with each other,
and exhibit characteristic structures e.g., flocking, vortex or seemingly chaotic behaviors, called the active turbulence.
Active matter is a typical example of non-equilibrium systems, and
unraveling the behavior of active systems contributes to the development of non-equilibrium physics.

Among various kinds of active systems,
active nematic \cite{doostmohammadi2018active} is a class of active matter which exhibits orientational order,
namely, constituents of active nematic have anisotropic shape and tend to align with each other,
just like nematic liquid crystals.
Examples of active nematic are
microtubule/motor protein mixtures,
bacterial suspensions,
cell populations,
etc \cite{doostmohammadi2018active}.

It is known that the behavior of active nematic can be described by the continuum theory
of nematic liquid crystals complemented by an active stress term \cite{doostmohammadi2018active}.
In the continuum theory of nematic systems,
the mean direction of constituents at a point $\bm{r}$ is described by a unit vector $\bm{n}\left(\bm{r}\right)$
called the director \cite{de1993physics,chandrasekhar1992liquid}.
The director does not have distinction between its head and tail, namely,
$\bm{n}\left(\bm{r}\right)$ and $-\bm{n}\left(\bm{r}\right)$ represent the identical state.
In the two-dimensional (2D) case,
which is exclusively discussed throughout this paper, the director can be identified by an angle $\Theta\left(\bm{r}\right)$
and written as $\bm{n}\left(\bm{r}\right)=\left(\cos\Theta\left(\bm{r}\right),~\sin\Theta\left(\bm{r}\right)\right)$.

A topological defect, which we may call simply a defect in the following,
is a singularity of the orientational field and can be classified by its charge.
The topological charge $q$ is defined
by $2\pi q=\int_C d\Theta=\int_C d\bm{r}\cdot\bm{\nabla}\Theta\left(\bm{r}\right)$,
where $C$ is a closed curve which encloses the defect \cite{kleman2003soft}.
As already mentioned, the director does not have the distinction between its head and tail, and then half-integer charges are allowed.
Under the one-constant approximation of the Frank free energy,
it can be shown that the energy of a topological defect of charge $q$ is proportional to $q^2$, and the interaction
force
between defects in 2D nematic
has the same form (proportional to $1/r$, where $r$ is the distance between defects)
as the 2D Coulomb interaction between electric charges \cite{de1993physics,chandrasekhar1992liquid}.
$\pm 1/2$ defects are energetically most stable, and frequently appear in nematic systems.

Topological defects are robust structures because they are topologically protected,
namely, they cannot be removed by a continuum deformation of the director field.
This robustness of topological defects is known to play an important role,
e.g., in morphogenesis \cite{maroudas2021topological}.
Understanding the behavior of topological defects under various conditions will
contribute to the development of not only active matter physics but also related fields, e.g., developmental biology.

Driven by the motivation mentioned above,
in this study, we consider the defect configuration of two dimensional active nematic around a circular obstacle.
In the case of passive nematic liquid crystals,
the equilibrium defect configuration around a circular obstacle can be identified by the method of image charge.
The analytical result agrees with the corresponding numerical calculation based on a continuum model \cite{fukuda2001director}
\footnote{Analytical and numerical calculations agree well when the elastic constant is sufficiently large
	and then the size of defect is sufficiently small.
		This is because, in the analytical calculation, we assume that the defect is point-like, namely, it has no spatial extent
}.
However, in the case of active nematic,
one has
to consider not only the Coulomb-like interaction between defects
but also forces exerted on defects by the fluid flow induced by the active stress.
In this situation, the analytical calculation considering only the Coulomb-like interaction is no longer valid.
Therefore,
we perform numerical simulations based on a continuum model.

Furthermore, to reveal the mechanism determining the defect configuration in our simulations,
we carry out a theoretical calculation
based on an analytical expression relating the defect velocity $\bm{v}$ with the profile of
the nematic order parameter and the flow field at the defect \cite{angheluta2021role,schimming2023kinematics,schimming2025analytical}.
By the condition $\bm{v}=0$, we derive a force balance equation of a defect and calculate the stationary position.
We find that the theoretical results qualitatively agree with the simulation results.

In this study, we focus on the low-activity regime, where
the orientation and the flow fields settle in a stationary profile.
The location of the topological defects in such a stationary state can be studied theoretically and numerically as mentioned above.
We do not investigate the high-activity regime where a number of topological defects spontaneously emerge and the system exhibits a turbulent state.
Although such a turbulent state has long been a subject of extensive studies, it is a challenging task to give an analytic argument on the dynamic arrangement of topological defects.
We stress that even the investigation of static states can give a deep insight into the  conditions  for the structural stability of an active system and the control of its morphology.

This paper is organized as follows:
In Sec.\ref{model}, we introduce the basic equations and the numerical method of our simulations.
We present our numerical and theoretical results and the discussion in Sec.\ref{result_discussion}.
Finally, we give conclusion and outlook in Sec.\ref{conclusion}.
 \section{Model and Numerical method}
\label{model}
\subsection{Basic equations}
\label{basic_equations}
The dynamics of two-dimensional active nematic is described by two variables,
the second-rank tensor order parameter $\bm{Q}$ and the velocity field $\bm{u}$.
The relation between the tensor order parameter $\bm{Q}$,
the scalar order parameter $S$ and the director $\bm{n}$ is given by $\bm{Q}=S(2\bm{n}\bm{n}-\bm{1})$, where $\bm{1}$ is the unit tensor.
The director $\bm{n}$ represents the average direction of the constituents,
and the scalar order parameter $S$ represents the degree of the orientational order.
To describe the behaviour of an active nematic,
we use a model consisting of the Edwards-Beris equation and the Stokes equation \cite{angheluta2021role}.
\newline
The Edwards-Beris equation gives the time evolution of the $\bm{Q}$-tensor \cite{angheluta2021role}:
\begin{equation}
	\begin{aligned}
		\label{EdwardsBerisEq}
		(\partial_t + \bm{u}\cdot\bm{\nabla})\bm{Q}
		=& \lambda\bm{E} + \bm{Q}\cdot\bm{\Omega} - \bm{\Omega}\cdot\bm{Q}\\
		&+ \frac{1}{\gamma}
		\left[
			K\bm{\nabla}^2\bm{Q} + g\left\{ 1 - {\rm{tr}}\left( \bm{Q}^2 \right)\right\}\bm{Q}
			\right].
	\end{aligned}
\end{equation}
where $\bm{E}=\frac{1}{2}\{\bm{\nabla}\bm{u}+(\bm{\nabla}\bm{u})^T-(\bm{\nabla}\cdot\bm{u})\bm{1}\}$ is the symmetric traceless part of the flow strain rate
and $\bm{\Omega}=\frac{1}{2}\{\bm{\nabla}\bm{u}-(\bm{\nabla}\bm{u})^T\}$ is the vorticity.
$\lambda$, $K$ and $\gamma$ are, respectively, the flow alignment parameter, the elastic constant and the rotational friction coefficient.
The last term $(g/\gamma)\left\{1-{\rm{tr}}(\bm{Q}^2)\right\}\bm{Q}$ drives
the system to the ordered state and $g$ is the strength of the ordering.
The perfect order corresponds to $S=1/\sqrt{2}\equiv S_{0}$ in our choice of coefficients.
\newline
The Stokes equation determines the velocity $\bm{u}$ for the given $\bm{Q}$-tensor \cite{angheluta2021role}:
\begin{equation}
	\begin{aligned}
		\label{StokesEq}
		0 = -\mu\bm{u}+\eta\bm{\nabla}^{2}\bm{u}+\alpha\bm{\nabla}\cdot\bm{Q},
	\end{aligned}
\end{equation}
The first term ($-\mu\bm{u}$) in the r.h.s. of eq.(\ref{StokesEq}) is the friction between active nematic and the substrate.
The second term ($\eta\bm{\nabla}^2\bm{u}$) is the shear viscosity,
and the third term ($\alpha\bm{\nabla}\cdot\bm{Q}$) represents the active stress.
The magnitude of $\alpha$ represents the strength of the activity.
$\alpha < 0$ and $\alpha > 0$ correspond to the extensile and contractile active nematic,
respectively \cite{doostmohammadi2018active}
\footnote{
	We note the difference between the definitions of the activity parameter in \cite{doostmohammadi2018active} and this paper.
	In \cite{doostmohammadi2018active}, the active stress tensor is defined as $-\zeta\bm{Q}$
	and thus $\zeta>0$ and $\zeta<0$ describe the extensile and contractile active nematic, respectively
}.
As in \cite{pokawanvit2022active,srivastava2016negative,angheluta2021role},
we use the compressible but constant density limit, where the density is assumed to be constant,
but $\bm{\nabla}\cdot\bm{u}=0$ is not enforced.
This means that the mass conservation is broken
but such a situation may be potentially realized by allowing the birth and death processes of constituents.
The pressure is assumed to depend only on the density, which is assumed to be constant,
and does not appear in the Stokes equation (Eq.(\ref{StokesEq})).
The simulations under the above-mentioned compressible but constant density limit
give qualitatively similar results to the ones obtained in particle-based simulations of
dry truly compressible fluids \cite{pokawanvit2022active}.
Furthermore, strictly,
the bulk viscosity term ($\propto\bm{\nabla}\left(\bm{\nabla}\cdot\bm{u}\right)$) should be included
in the Stokes equation (Eq.(\ref{StokesEq}))
since we do not enforce $\bm{\nabla}\cdot\bm{u}=0$.
However, this term will make our analysis of the Stokes equation
(see Sec.\ref{analytical_expression_for_the_flow_field} or Appendix \ref{review_on_flow_field}) quite complicated.
Thus, we ignore this term in this paper.

Let us nondimensionalize the basic equations (eqs.(\ref{EdwardsBerisEq}) and (\ref{StokesEq}))
using the nematic correlation length $l\equiv\sqrt{K/g}$ and the nematic relaxation time $\tau\equiv\gamma/g$.
After the nondimensionalization, our basic equation reduces to
\begin{equation}
	\begin{aligned}
		\label{EdwardsBerisEq_nondim}
		(\partial_{\tilde{t}} + \tilde{\bm{u}}\cdot\tilde{\bm{\nabla}})\bm{Q}
		=& \lambda\tilde{\bm{E}} + \bm{Q}\cdot\tilde{\bm\Omega} - \tilde{\bm\Omega}\cdot\bm{Q}\\
		&+ \tilde{\bm\nabla}^{2}\bm{Q} + \left\{ 1 - {\rm{tr}}\left( \bm{Q}^2 \right)\right\}\bm{Q}
	\end{aligned}
\end{equation}
and
\begin{equation}
	\begin{aligned}
		\label{StokesEq_nondim}
		0 = - \tilde{\bm u}+ \tilde{\eta}\tilde{\bm\nabla}^{2}\tilde{\bm u}+\tilde{\alpha}\tilde{\bm\nabla}\cdot\bm{Q},
	\end{aligned}
\end{equation}
where we have denoted nondimensionalized quantities and parameters by symbols with tilde.
Relations between dimensional and nondimensional quantities are listed in TABLE.\ref{relation_dim_and_nondim}.
\begin{table}[btp]
	\begin{center}
		\caption{
			Relations between dimensional and nondimensional quantities
		}
		\label{relation_dim_and_nondim}
\begin{tabular}{c|c|c}
			\begin{tabular}{c}Nondimensional\\quantity\end{tabular} & Symbol & \begin{tabular}{c}In terms of\\dimensional quantities\end{tabular}\\ \hline
				elastic constant           & $\tilde{K}$            & ${K}/(gl^2)={K}/(g\sqrt{K/g}^2)=1$\\
				viscosity                  & $\tilde{\eta}$         & ${\eta}/(\mu l^2)$\\
				activity                   & $\tilde{\alpha}$       & ${\alpha \tau}/(\mu l^2)$\\
				time derivative            & $\partial_{\tilde{t}}$ & $\tau\partial_t$\\
				fluid velocity             & $\tilde{\bm{u}}$       & $\tau\bm{u}/l$\\
				nabla                      & $\tilde{\bm{\nabla}}$  & $l\bm{\nabla}$\\
				\begin{tabular}{c}symmetric traceless\\part of the flow\\strain rate\end{tabular}
					& $\tilde{\bm{E}}$       & $\tau\bm{E}$\\
					vorticity tensor  & $\tilde{\bm{\Omega}}$  & $\tau\bm{\Omega}$\\ \hline
		\end{tabular}
\end{center}
\end{table}
For simplicity, we will omit tilde in the following.

\subsection{Numerical method}
\label{numerical_method}
We use the two dimensional polar coordinate $(r,\theta)$ for our numerical calculations.
The tensor order parameter $\bm{Q}$ has two degrees of freedom specified by $S$ and the director of $\bm{n}$.
Let us choose $Q_{rr}$ and $Q_{r\theta}$ as two independent components.
For our numerical simulations,
we first rewrite the basic equations (eqs.(\ref{EdwardsBerisEq_nondim}) and (\ref{StokesEq_nondim}))
in terms of $Q_{rr}$, $Q_{r\theta}$, $u_{r}$ and $u_{\theta}$ in the $(r,\theta)$-coordinates.
Furthermore,
we introduce an additional coordinate $\xi\equiv\ln(r/R_{0})$, where $R_{0}$ is the radius of the circular obstacle
and equal grid spacings are taken in the $(\xi,\theta)$ plane.
Thus, in the real space, the lattice spacing becomes larger away from the obstacle.
This can be justified because we are interested in the behaviour near the obstacle.
\newline
\indent
After a straightforward calculation (see Appendix.\ref{equations_in_xi_theta_coordinate}),
we can obtain our basic equation.

To calculate the time evolution of $Q_{rr}$ and $Q_{r\theta}$, we use the fourth order Runge-Kutta formula.
The Stokes equation is solved by a solver for linear algebra problems \cite{cpp_eigen}.
For the detailed procedure, see Appendix.\ref{how_to_solve_Stokes_eq}.
To reduce computation time,
we solve the Stokes equation every 1000 time steps.
This can be justified because we are interested in the final stationary state.

Parameters in our model are $\lambda$, $\eta$, $\alpha$ and $R_{0}$.
Numerical values of the parameters used in simulations are listed in TABLE.\ref{parameter_values}.
\begin{table}[btp]
	\begin{center}
		\caption{
			Numerical values of parameters in simulations.
		}
		\label{parameter_values}
\begin{tabular}{c|c|c}
			Parameter & Symbol   & Numerical value\\ \hline
			flow alignment parameter & $\lambda$ & 1.5\\
			viscosity               & $\eta$    & $10^2$\\
			activity                & $\alpha$  & $-1.0$\\
			&           & $-0.8$\\
			&           & $-0.6$\\
			&           & $-0.4$\\
			&           & $-0.2$\\
			&           & $0$\\
			&           & $0.2$\\
			&           & $0.4$\\
			&           & $0.6$\\
			&           & $0.8$\\
			&           & $1.0$\\
			Radius of obstacle      & $R_{0}$   & $10^{2}$\\
			\begin{tabular}{c}
				Number of lattice points\\in the $\xi$-direction 
			\end{tabular}
			& $N$ & $480$\\
			\begin{tabular}{c}
				Number of lattice points\\in the $\theta$-direction
			\end{tabular}
			& $M$ & $1000$\\
			Lattice spacing in the $\xi$-direction & $\Delta\xi$ & $\ln\left(1+\Delta\theta\right)$\\
			Lattice spacing in the $\theta$-direction & $\Delta\theta$ & $2\pi/M$\\
			Time increment & $\Delta t$ & $10^{-1}$\\
		\end{tabular}
\end{center}
\end{table}
 \subsection{Setup}
\label{setup}
\begin{figure*}[btp]
	\begin{minipage}[b]{0.49\linewidth}
		\centering
		\includegraphics[width=0.95\linewidth]{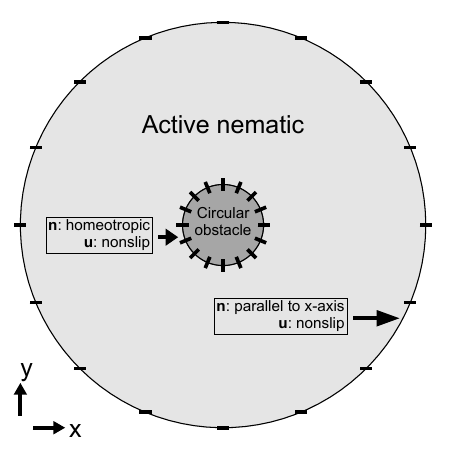}
		\subcaption{Homeotropic}
	\end{minipage}
	\begin{minipage}[b]{0.49\linewidth}
		\centering
		\includegraphics[width=0.95\linewidth]{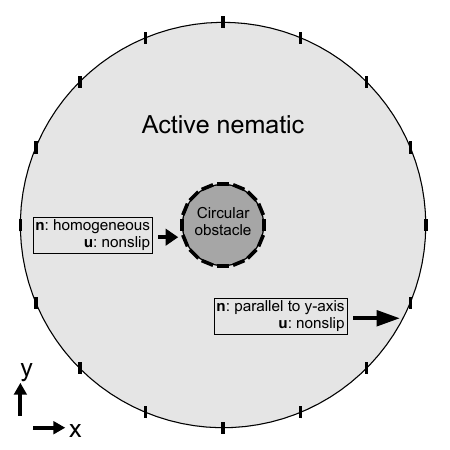}
		\subcaption{Homogeneous}
	\end{minipage}
	\caption{
		The schematic of the setup of our simulations.
		We carry out simulations under two different boundary conditions on $\bm{n}$.
		We adopt
		(a) the homeotropic condition ($\bm{n}$ is perpendicular to the surface of the obstacle),
		and
		(b) the homogeneous condition ($\bm{n}$ is parallel to the surface of the obstacle).
		Black bars at inner and outer boundaries represent the director.
		The light gray region is filled with active nematic and the dark gray circle represents the circular obstacle.
		Note that the ratio of the radius of the obstacle to the radius of the outer boundaries
		is not identical with the one in our simulations.
	}
	\label{setup_fig}
\end{figure*}

As already mentioned in Sec.\ref{introduction},
we are interested in the behaviour of active nematic around a circular obstacle.
Our setup of numerical simulations is schematically described in FIG.\ref{setup_fig}.
A circular obstacle is located in active nematic
and outer boundary is set sufficiency far from the obstacle because our interest is in the director configuration near the obstacle.
We adopt two types of boundary conditions on $\bm{n}$, the homeotropic and the homogeneous ones.
In the homeotropic (resp. homogeneous) boundary condition, $\bm{n}$ is perpendicular (resp. parallel) to the surface of the obstacle.
At the outer boundary, we set $\bm{n}$ parallel to the $x$-axis (resp. $y$-axis) in the homeotropic (resp. homogeneous) case (see FIG.\ref{setup_fig}),
so that two $-1/2$ defects (see Sec.\ref{result_discussion}) emerge on the $y$-axis in both cases.
The scalar order parameter $S$ is fixed at $S_{0}=1/\sqrt{2}$ and
the velocity field is fixed at $0$ (the nonslip boundary condition) at both boundaries.

The initial condition is set as follows:
We first perform the simulation without activity ($\alpha$=0), and obtain the equilibrium profile of $\bm{Q}$
(see Sec.\ref{result_discussion_passive_case}).
We adopt the equilibrium profile as the initial condition of $\bm{Q}$ in the active case,
and the initial velocity field is calculated from the initial profile of $\bm{Q}$ via the Stokes equation (eq.(\ref{StokesEq_nondim})).

The initial condition of the simulations with $\alpha=0$ is given as follows:
$\bm{n}$ and $S$ at the boundaries are set as stated above.
In the bulk, $\bm{n}$ is set to uniform profile parallel to the $x$-axis (resp. $y$-axis)
with small perturbations in the homeotropic (resp. homogeneous) case, and $S=0.7071$.
 \section{Results and Discussion}
\label{result_discussion}
\subsection{Results of $\alpha=0$ case (Passive nematic)}
\label{result_discussion_passive_case}
Before discussing the simulations of active nematic, let us review the case of (passive) nematic liquid crystals.
When $\alpha=0$, Eq.(\ref{StokesEq}) reads $0=-\bm{u}+\eta\bm{\nabla}^2\bm{u}$.
This equation is composed of only the friction and viscosity term, and then the solution is $\bm{u}=0$
under the nonslip boundary condition at both boundaries.
In this situation, the evolution equation for $\bm{Q}$-tensor reduces to
\begin{equation}
	\label{model_A}
	\partial_t \bm{Q}
	= -\frac{\delta\mathcal{F}}{\delta\bm{Q}}
	=\bm{\nabla}^2\bm{Q}+\left\{1-{\rm{tr}}\left(\bm{Q}^2\right)\right\}\bm{Q},
\end{equation}
where
\begin{equation}
	\label{model_A_functional}
	\mathcal{F}=\int d\bm{r}\left[
		\frac{1}{2} \left(\partial_i Q_{jk}\right) \left(\partial_i Q_{jk}\right)
		+ \frac{1}{4}\left\{
			1 - {\rm{tr}}\left(\bm{Q}^2\right)
			\right\}^{2}
		\right].
\end{equation}
Eq.(\ref{model_A}) is the relaxation dynamics of model A \cite{hohenberg1977theory}.
The equilibrium profile of $\bm{Q}$ minimizes the free energy $\mathcal{F}$.

In the limit of $R_{0}\gg 1$, namely, the size of the obstacle is sufficiently larger than the nematic coherence length
(=1, in our choice of the unit length), one can regard a defect as a point charge.
In this situation, the configuration of topological defects around a circular obstacle
can be calculated based on the Coulomb-like interaction between defects and the method of image charge.

At the surface of the obstacle,
the homeotropic or homogeneous boundary condition is imposed,
and then the obstacle has
a $+1$ charge.
At the outer boundary,
the director is parallel to the $x$-axis or $y$-axis everywhere,
and then the total charge of this system is $0$.
Therefore,
the total charge of topological defects in the liquid crystal is $-1$,
and two $-1/2$ defects will emerge because they are more stable than one $-1$ defect.

\begin{figure*}[btp]
	\begin{minipage}[b]{0.49\linewidth}
		\centering
		\includegraphics[width=0.99\linewidth]{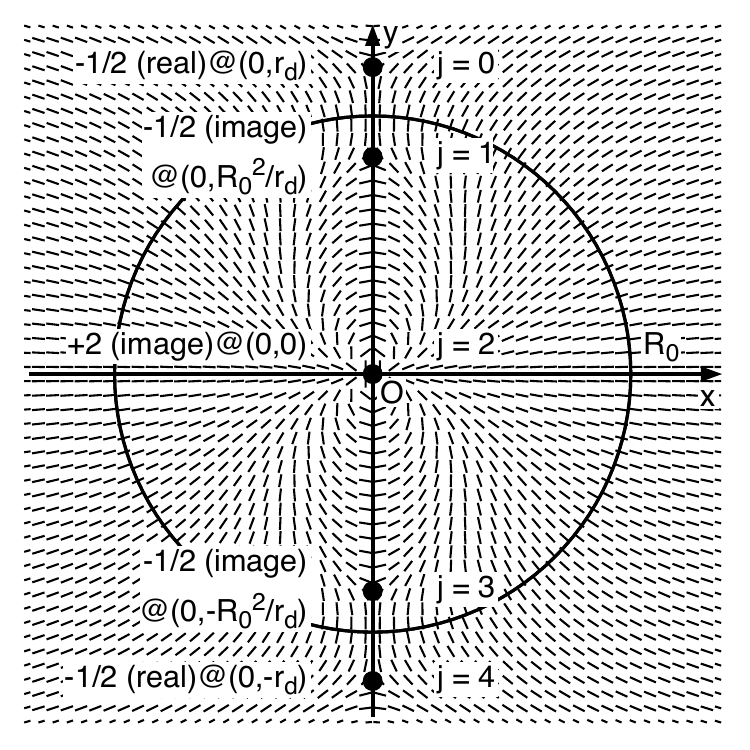}
		\subcaption{Homeotropic}
	\end{minipage}
	\begin{minipage}[b]{0.49\linewidth}
		\centering
		\includegraphics[width=0.99\linewidth]{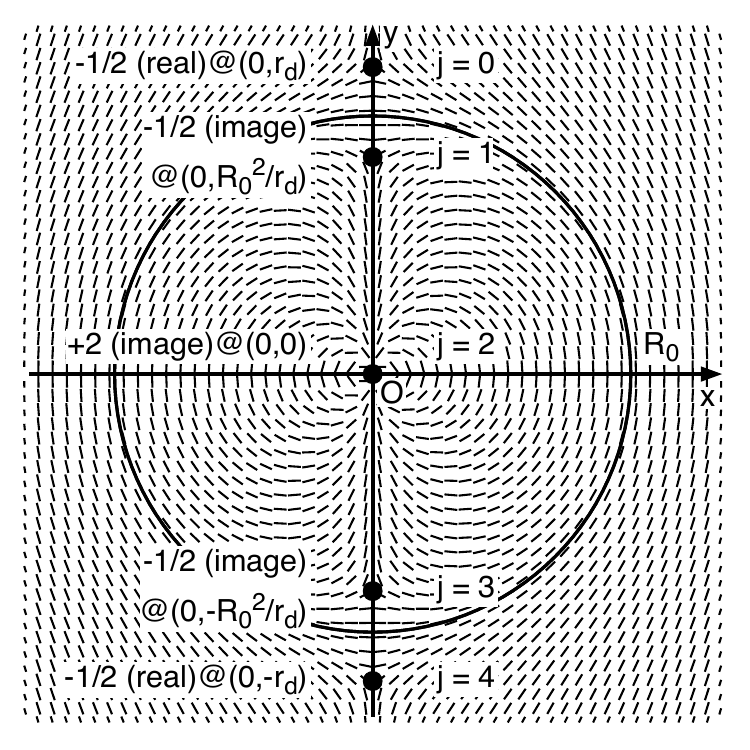}
		\subcaption{Homogeneous}
	\end{minipage}
	\caption{
		The defect configuration satisfying
		(a) the homeotropic and (b) the homogeneous boundary conditions at the surface of the obstacle.
		$\bm{n}$ is parallel to (a) $x$-axis and (b) $y$-axis in the limit of $r=\sqrt{x^{2}+y^{2}}\rightarrow\infty$.
		The circle indicates the surface of the obstacle, and $R_{0}$ is its radius.
		The bars represent the director field.
		Filled circles on the $y$-axis indicate positions of defects.
		$j$ is the label for defects introduced in Sec.\ref{the_halperin_mazenko_formalism}.
	}
	\label{configuration_of_image_charges}
\end{figure*}

While the two $-1/2$ defects repel each other, they are attracted by the $+1$ charge of the obstacle.
When those opposing forces balance, the $-1/2$ defects will stabilize,
and be located on the $y$-axis.
We denote the distance between the origin and one $-1/2$ defect by $r_{d}$ (see FIG.\ref{configuration_of_image_charges}).

The homeotropic or homogeneous boundary condition can be satisfied
by introducing imaginary defects $-1/2$, $+2$ and $-1/2$ at $(0,+R_{0}^2/r_{d})$,
$(0,0)$ and $(0,-R_{0}^2/r_{d})$, respectively (see FIG.\ref{configuration_of_image_charges}) \cite{fukuda2001director}.
 From the symmetry, it is sufficient to consider the force balance for one of the two real defects.
The force balance equation for one real $-1/2$ defect is given by
\begin{equation}
	\frac{-\frac{1}{2}}{r_{d}-\frac{R_{0}^2}{r_{d}}}
	+\frac{+2}{r_{d}}
	+\frac{-\frac{1}{2}}{r_{d}+\frac{R_{0}^2}{r_{d}}}
	+\frac{-\frac{1}{2}}{2r_{d}}=0,
\end{equation}
which can be solved easily and gives $r_{d}=(7/3)^{1/4}R_{0}$.

\begin{figure*}[btp]
	\begin{minipage}[b]{0.49\linewidth}
		\centering
		\includegraphics[width=0.99\linewidth]{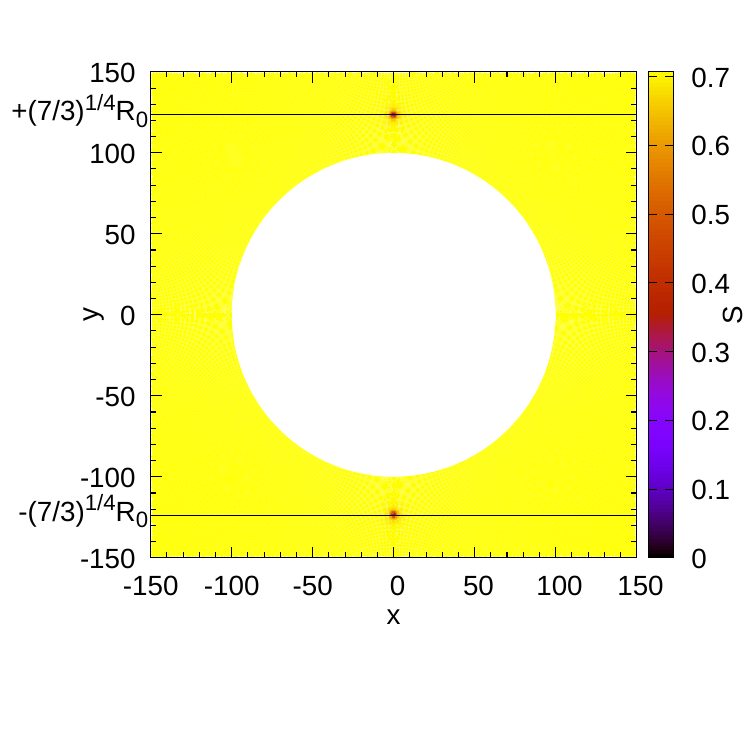}
		\subcaption{
			Scalar order parameter, Homeotropic.
		}
	\end{minipage}
	\begin{minipage}[b]{0.49\linewidth}
		\centering
		\includegraphics[width=0.99\linewidth]{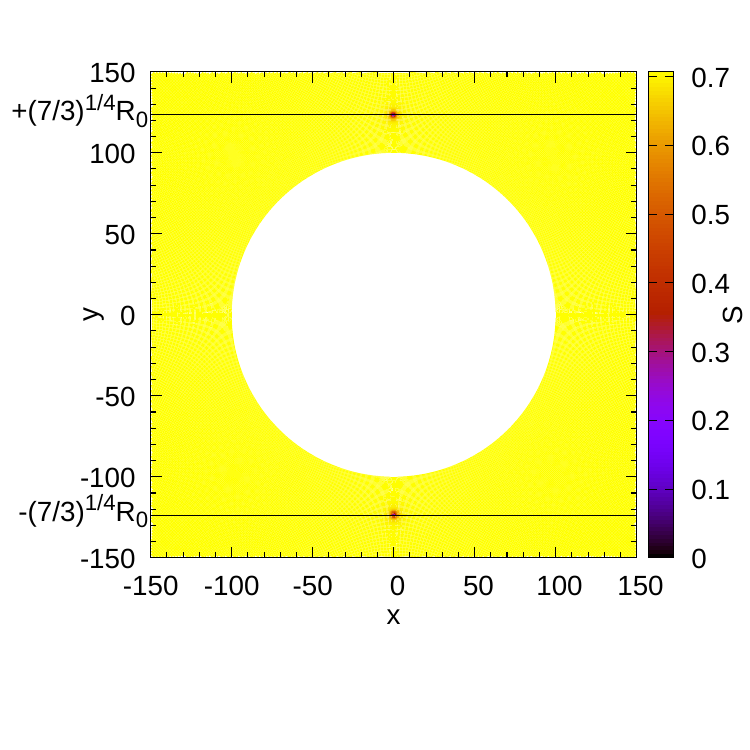}
		\subcaption{
			Scalar order parameter, Homogeneous.
		}
	\end{minipage}\\
\begin{minipage}[b]{0.49\linewidth}
		\centering
		\includegraphics[width=0.99\linewidth]{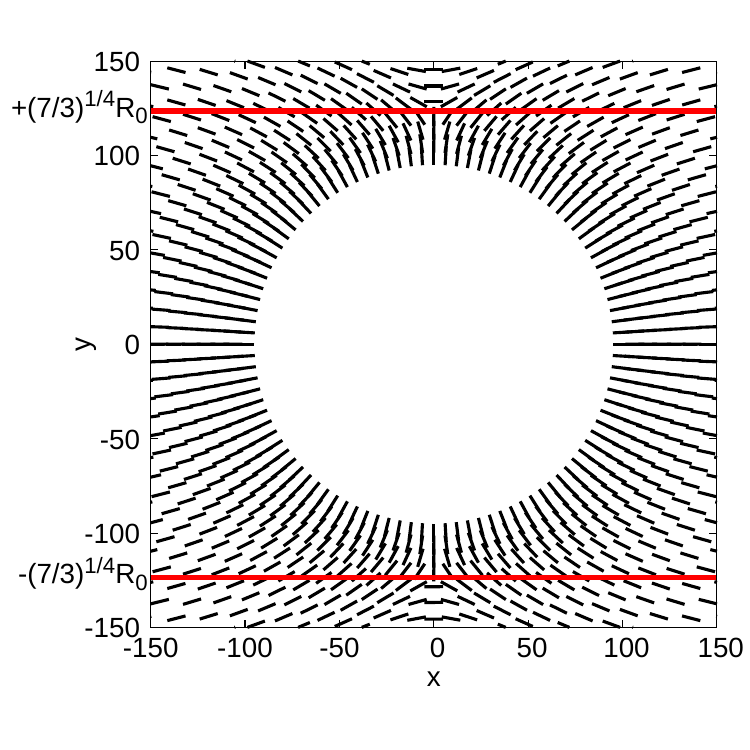}
		\subcaption{
			Director, Homeotropic.
		}
	\end{minipage}
	\begin{minipage}[b]{0.49\linewidth}
		\centering
		\includegraphics[width=0.99\linewidth]{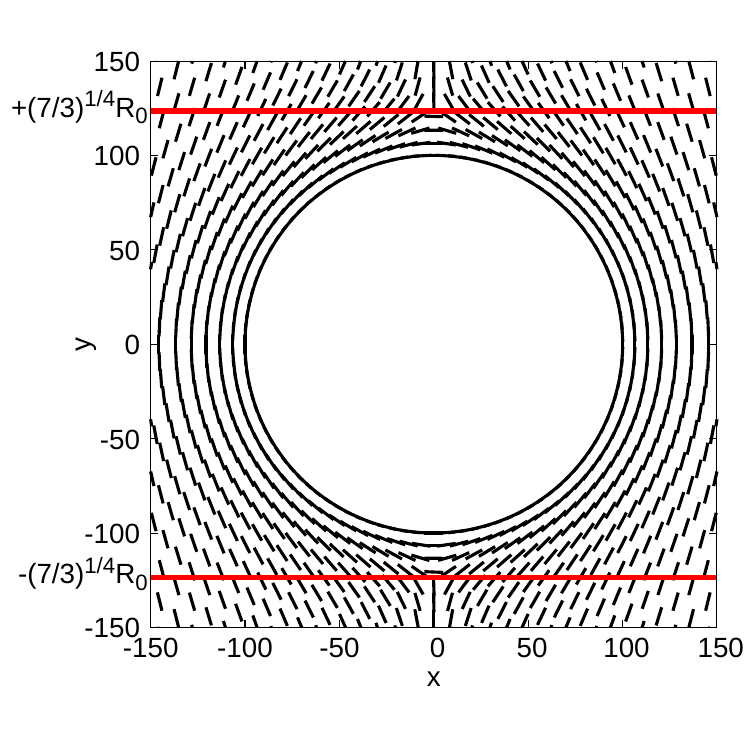}
		\subcaption{
			Director, Homogeneous.
		}
	\end{minipage}
	\caption{
		The simulation results of the passive case ($\alpha=0$) under the
		(a, c) homeotropic
		and the
		(b, d) homogeneous boundary conditions.
		(a, b)
		and
		(c, d)
		are the stationary profiles (at $t=20000$) of the scalar order parameter $S$ and the director $\bm{n}$, respectively.
		The horizontal lines indicate the equilibrium positions of defects predicted by the analytical calculation.
		The white circular region represents the obstacle.
	}
	\label{result_passive}
\end{figure*}

The above analytical result can be confirmed by the numerical simulation, where we calculate the evolution equation (Eq.(\ref{model_A}))
until the profile settles in a stationary state.
The results of numerical calculation are shown in FIG.\ref{result_passive}.
We can confirm that the analytical calculation correctly predicts the defect positions in the numerical simulation.

In the next section, we perform numerical simulations for $\alpha\neq 0$ cases,
and investigate how the defects configuration deviates from the passive case.
 \subsection{Results of $\alpha\neq 0$ case (Active nematic)}
\label{result_discussion_active_case}
Unlike in the case of $\alpha = 0$ (Sec.\ref{result_discussion_passive_case}),
when $\alpha\neq 0$, the r.h.s. of Eq.(\ref{EdwardsBerisEq_nondim}) cannot be written
by the functional derivative of some functional alone, and then it is not trivial whether the profile of $\bm{Q}$ and $\bm{u}$ settles in a stationary state.
However, we confirmed numerically that it is the case for low activity (small-$|\alpha|$) or high viscosity (large-$\eta$).
For large-$|\alpha|$ or small-$\eta$, the profile of
$\bm{Q}$
is no longer stable and becomes
turbulent.
Such a turbulent state around a circular obstacle will be a subject of future study, and
we focus only on the stationary profile in this paper.

\begin{figure*}[btp]
\begin{minipage}[b]{0.32\linewidth}
		\centering
		\includegraphics[width=0.99\linewidth]{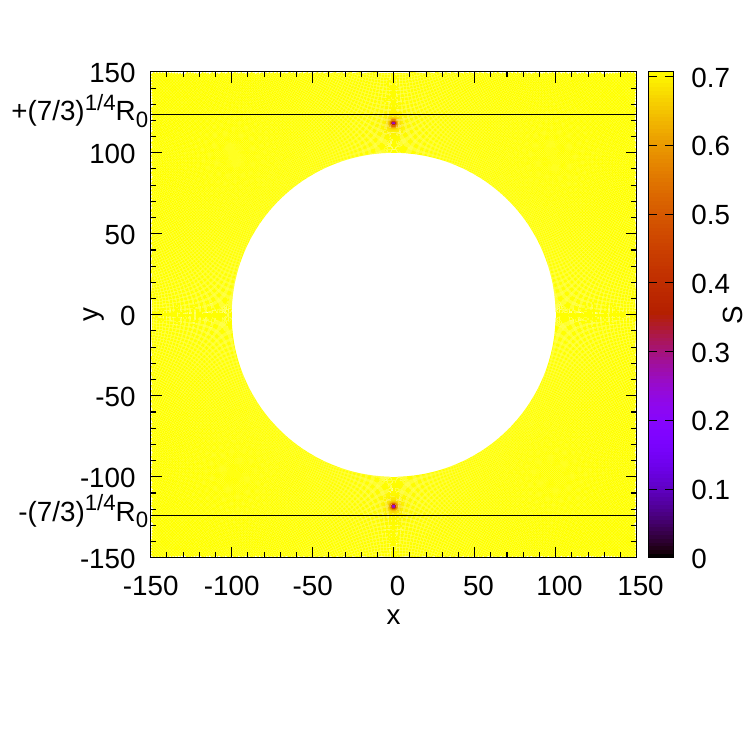}
		\subcaption{
			Scalar order parameter.
			Homeotropic.
			$\alpha=0.6$.
		}
		\label{fig_scalar_order_parameter_active_homeotropic_positive_alpha}
	\end{minipage}
	\begin{minipage}[b]{0.32\linewidth}
		\centering
		\includegraphics[width=0.99\linewidth]{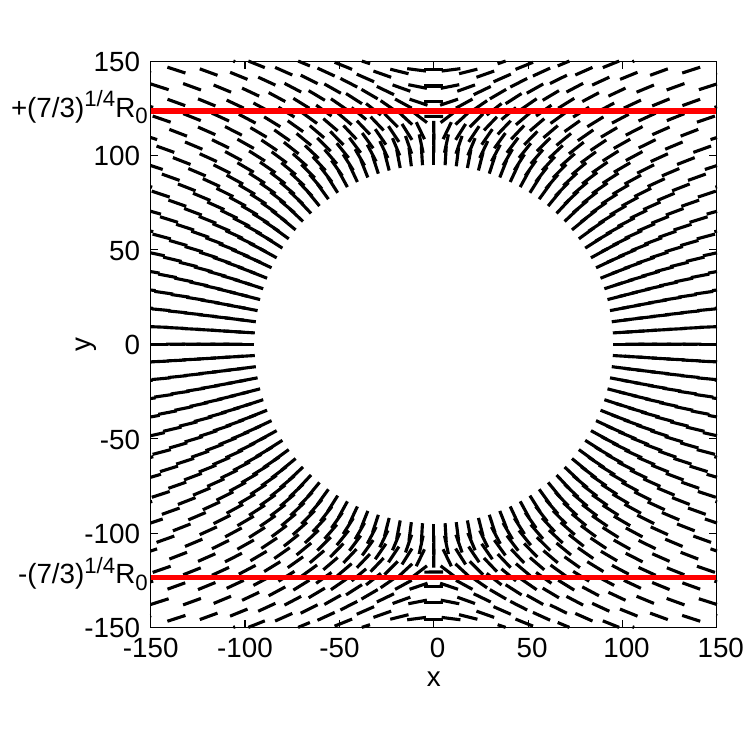}
		\subcaption{
			Director.
			Homeotropic.
			$\alpha=0.6$.
		}
		\label{fig_director_active_homeotropic_positive_alpha}
	\end{minipage}
	\begin{minipage}[b]{0.32\linewidth}
		\centering
		\includegraphics[width=0.99\linewidth]{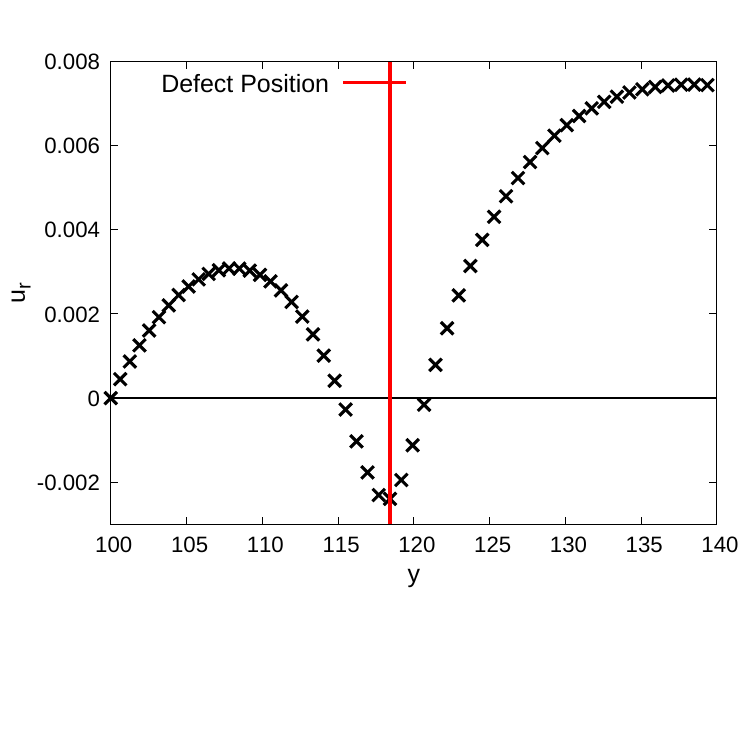}
		\subcaption{
			Velocity field on $y$-axis.
			Homeotropic.
			$\alpha=0.6$.
		}
		\label{fig_velocity_on_y_axis_active_homeotropic_positive_alpha}
	\end{minipage}
\begin{minipage}[b]{0.32\linewidth}
		\centering
		\includegraphics[width=0.99\linewidth]{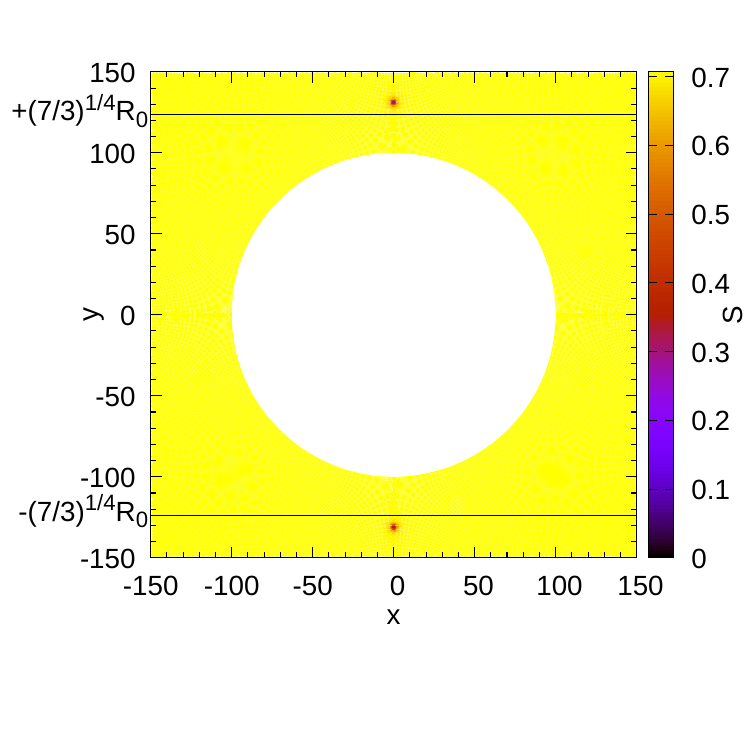}
		\subcaption{
			Scalar order parameter.
			Homeotropic.
			$\alpha=-0.6$.
		}
		\label{fig_scalar_order_parameter_active_homeotropic_negative_alpha}
	\end{minipage}
	\begin{minipage}[b]{0.32\linewidth}
		\centering
		\includegraphics[width=0.99\linewidth]{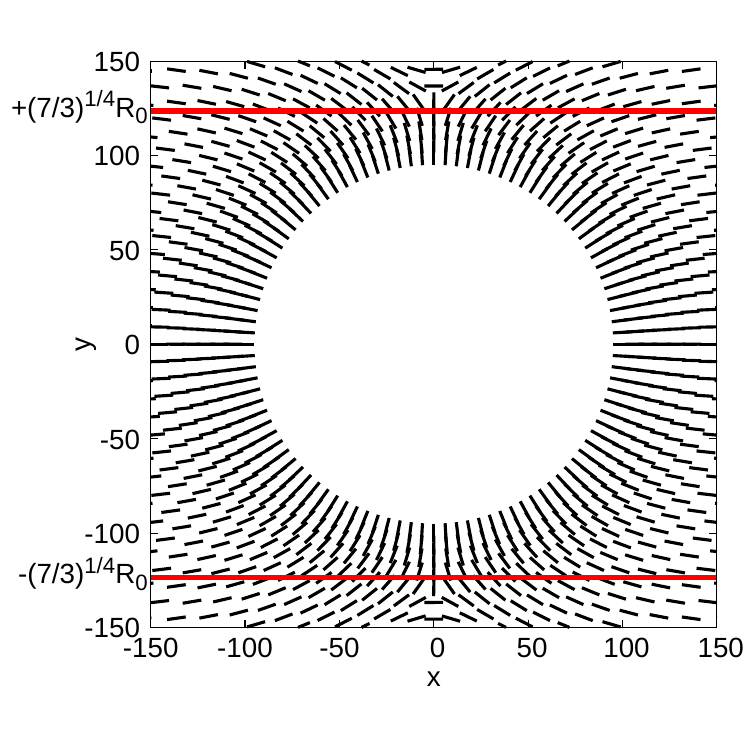}
		\subcaption{
			Director.
			Homeotropic.
			$\alpha=-0.6$.
		}
		\label{fig_director_active_homeotropic_negative_alpha}
	\end{minipage}
	\begin{minipage}[b]{0.32\linewidth}
		\centering
		\includegraphics[width=0.99\linewidth]{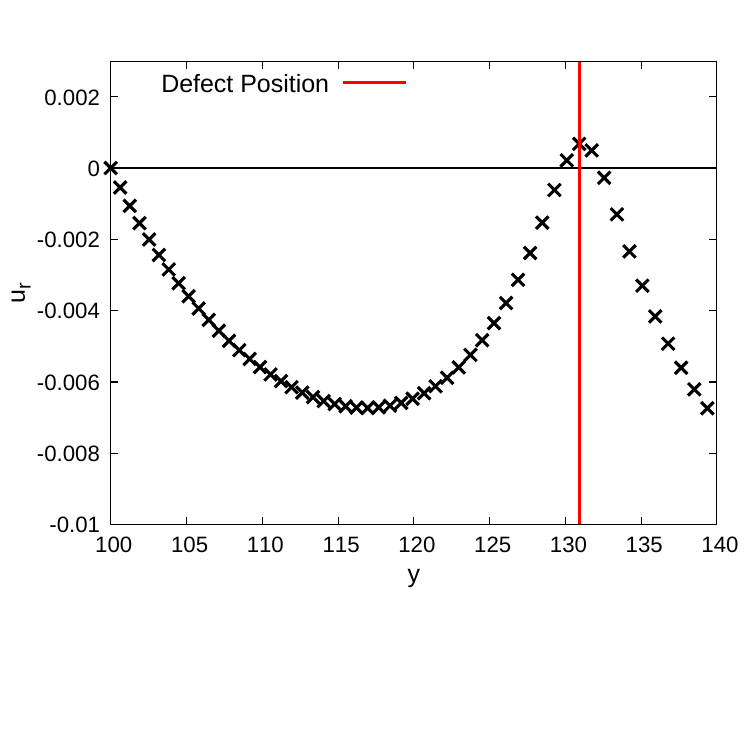}
		\subcaption{
			Velocity field on $y$-axis.
			Homeotropic.
			$\alpha=-0.6$.
		}
		\label{fig_velocity_on_y_axis_active_homeotropic_negative_alpha}
	\end{minipage}
	\caption{
		The typical results of simulations in active cases under the homeotropic boundary condition at $t=30000$.
		(a), (b) and (c) are the results of simulations with $\alpha = 0.6$ (a contractile case), and
		(d), (e) and (f) are the ones with $\alpha = -0.6$ (an extensile case).
		(a) and (d) are the scalar order parameter fields.
		(b) and (e) are the director fields.
		(c) and (f) represent the dependence of the radial component of the velocity $u_{r}$ on the $y$-coordinate.
	}
	\label{result_active_homeotropic_case}
\end{figure*}

\begin{figure*}[btp]
\begin{minipage}[b]{0.32\linewidth}
		\centering
		\includegraphics[width=0.99\linewidth]{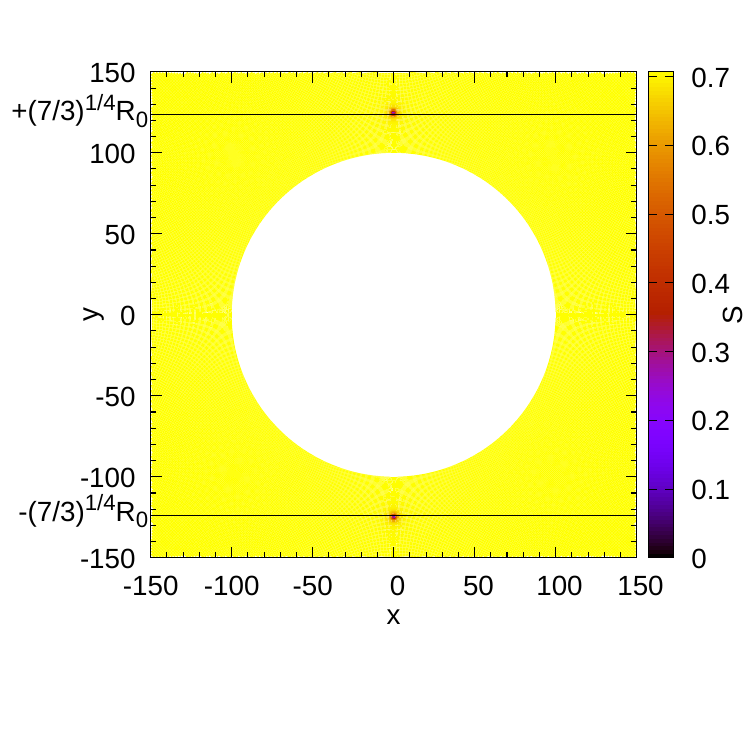}
		\subcaption{
			Scalar order parameter.
			Homogeneous.
			$\alpha=0.6$.
		}
		\label{fig_scalar_order_parameter_active_homogeneous_positive_alpha}
	\end{minipage}
	\begin{minipage}[b]{0.32\linewidth}
		\centering
		\includegraphics[width=0.99\linewidth]{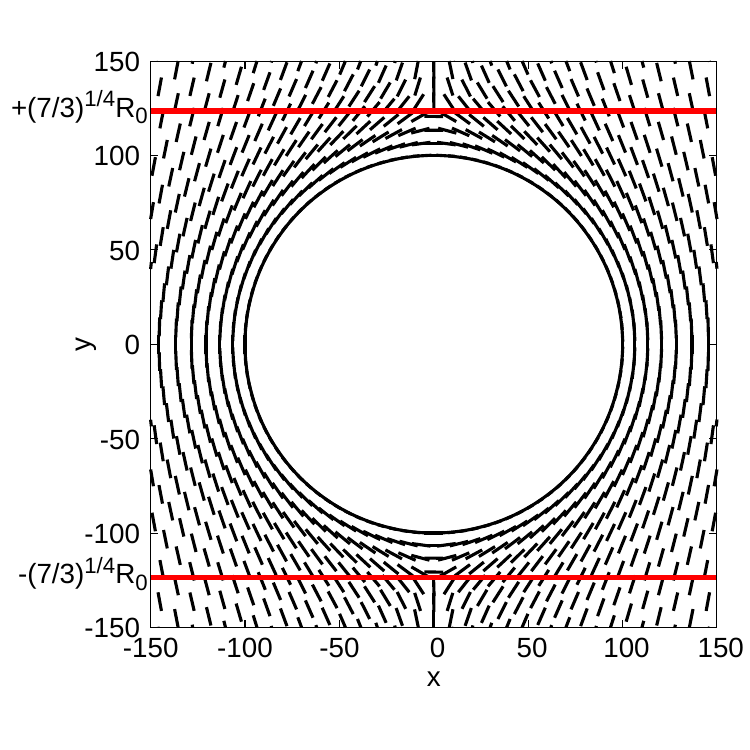}
		\subcaption{
			Director.
			Homogeneous.
			$\alpha=0.6$.
		}
		\label{fig_director_active_homogeneous_positive_alpha}
	\end{minipage}
	\begin{minipage}[b]{0.32\linewidth}
		\centering
		\includegraphics[width=0.99\linewidth]{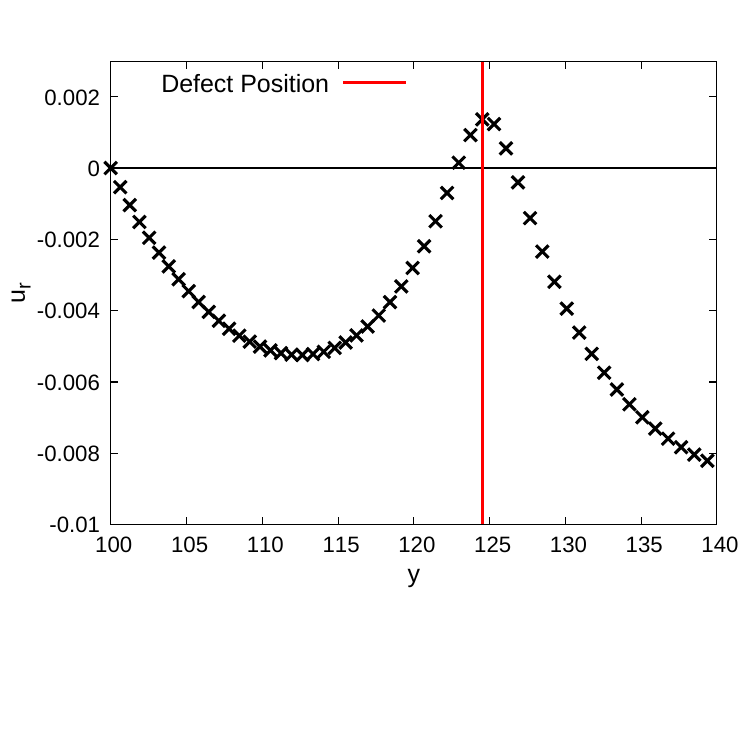}
		\subcaption{
			Velocity field on $y$-axis.
			Homogeneous.
			$\alpha=0.6$.
		}
		\label{fig_velocity_on_y_axis_active_homogeneous_positive_alpha}
	\end{minipage}
\begin{minipage}[b]{0.32\linewidth}
		\centering
		\includegraphics[width=0.99\linewidth]{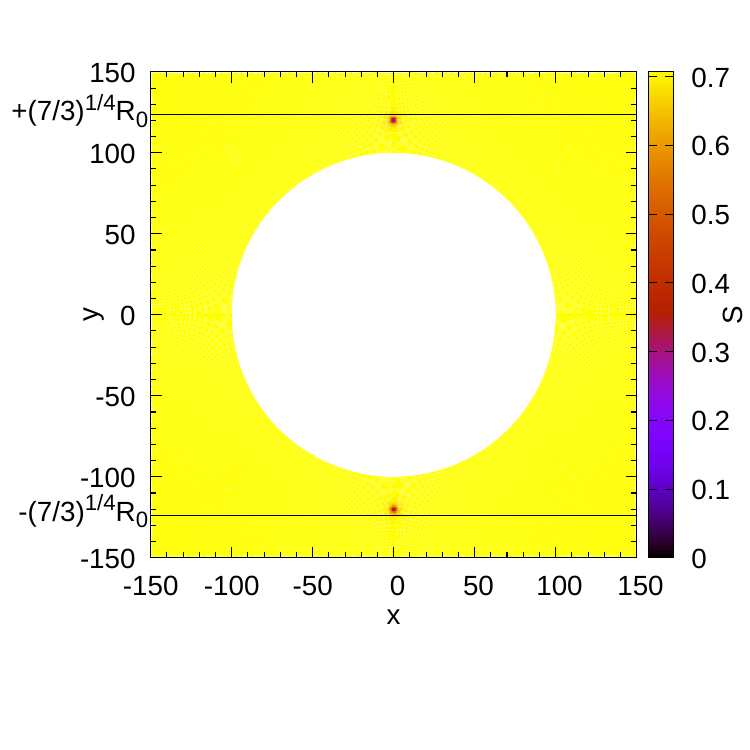}
		\subcaption{
			Scalar order parameter.
			Homogeneous.
			$\alpha=-0.6$.
		}
		\label{fig_scalar_order_parameter_active_homogeneous_negative_alpha}
	\end{minipage}
	\begin{minipage}[b]{0.32\linewidth}
		\centering
		\includegraphics[width=0.99\linewidth]{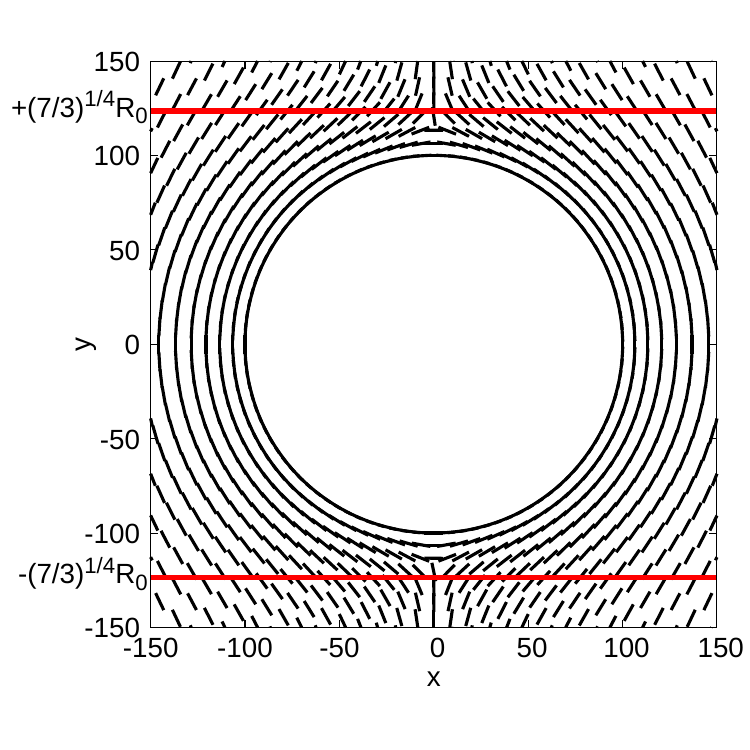}
		\subcaption{
			Director.
			Homogeneous.
			$\alpha=-0.6$.
		}
		\label{fig_director_active_homogeneous_negative_alpha}
	\end{minipage}
	\begin{minipage}[b]{0.32\linewidth}
		\centering
		\includegraphics[width=0.99\linewidth]{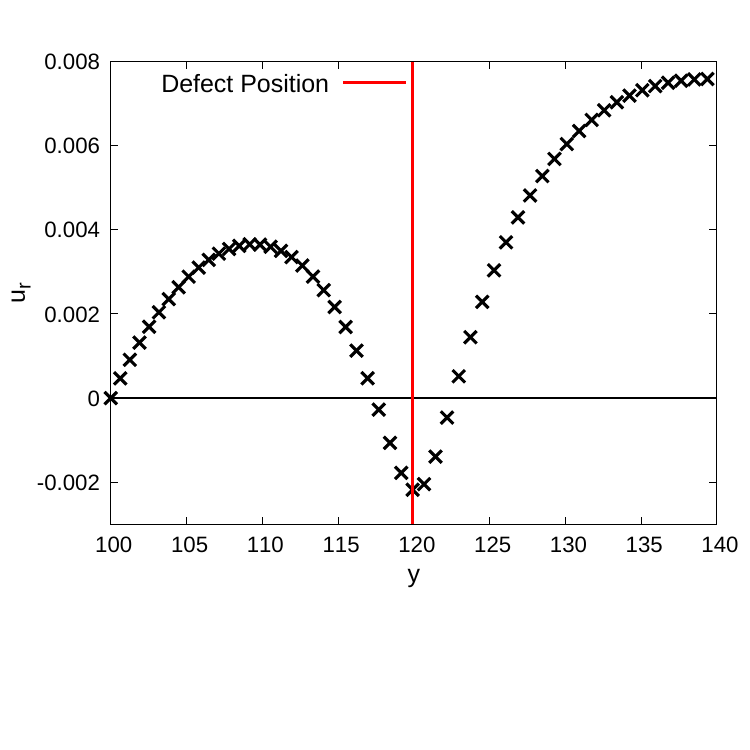}
		\subcaption{
			Velocity field on $y$-axis.
			Homogeneous.
			$\alpha=-0.6$.
		}
		\label{fig_velocity_on_y_axis_active_homogeneous_negative_alpha}
	\end{minipage}
	\caption{
		The typical results of simulations in active cases under the homogeneous boundary condition at $t=30000$.
		(a), (b) and (c) are the results of simulations with $\alpha = 0.6$ (a contractile case), and
		(d), (e) and (f) are the ones with $\alpha = -0.6$ (an extensile case).
		(a) and (d) are the scalar order parameter fields.
		(b) and (e) are the director fields.
		(c) and (f) represent the dependence of the radial component of the velocity $u_{r}$ on the $y$-coordinate.
	}
	\label{result_active_homogeneous_case}
\end{figure*}

Examples of simulation results under the homeotropic boundary condition are shown in FIG.\ref{result_active_homeotropic_case}.
It shows that positions of $-1/2$ defects shift toward (resp. away from) the obstacle in the contractile
(see FIG.\ref{fig_scalar_order_parameter_active_homeotropic_positive_alpha},
\ref{fig_director_active_homeotropic_positive_alpha},
\ref{fig_velocity_on_y_axis_active_homeotropic_positive_alpha})
(resp. extensile
(see FIG.\ref{fig_scalar_order_parameter_active_homeotropic_negative_alpha},
\ref{fig_director_active_homeotropic_negative_alpha},
\ref{fig_velocity_on_y_axis_active_homeotropic_negative_alpha})
) case.
FIG.\ref{result_active_homogeneous_case} shows examples of simulation results under the homogeneous boundary condition.
We can observe the opposite tendency from the homeotropic case.
However, the deviation in the contractile case is quite small compared to the homeotropic case.

\begin{figure*}[btp]
	\begin{minipage}[b]{0.49\linewidth}
		\centering
		\includegraphics[width=0.99\linewidth]{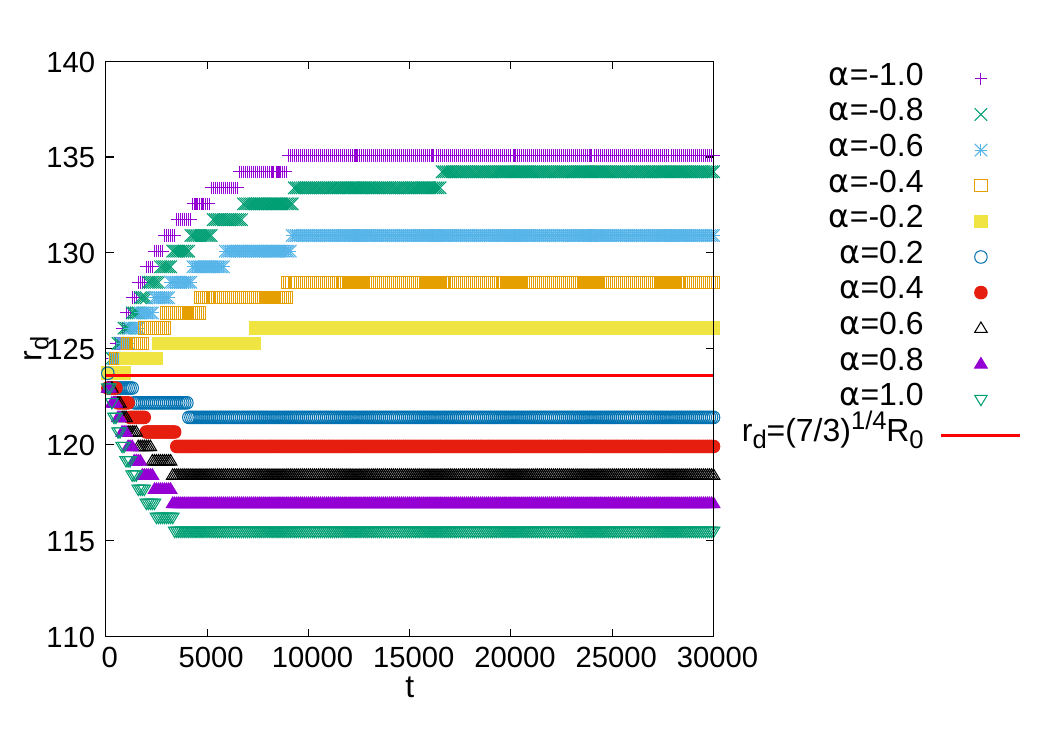}
		\subcaption{
			Homeotropic.
		}
		\label{fig_time_evolution_of_defect_position_homeotropic}
	\end{minipage}
	\begin{minipage}[b]{0.49\linewidth}
		\centering
		\includegraphics[width=0.99\linewidth]{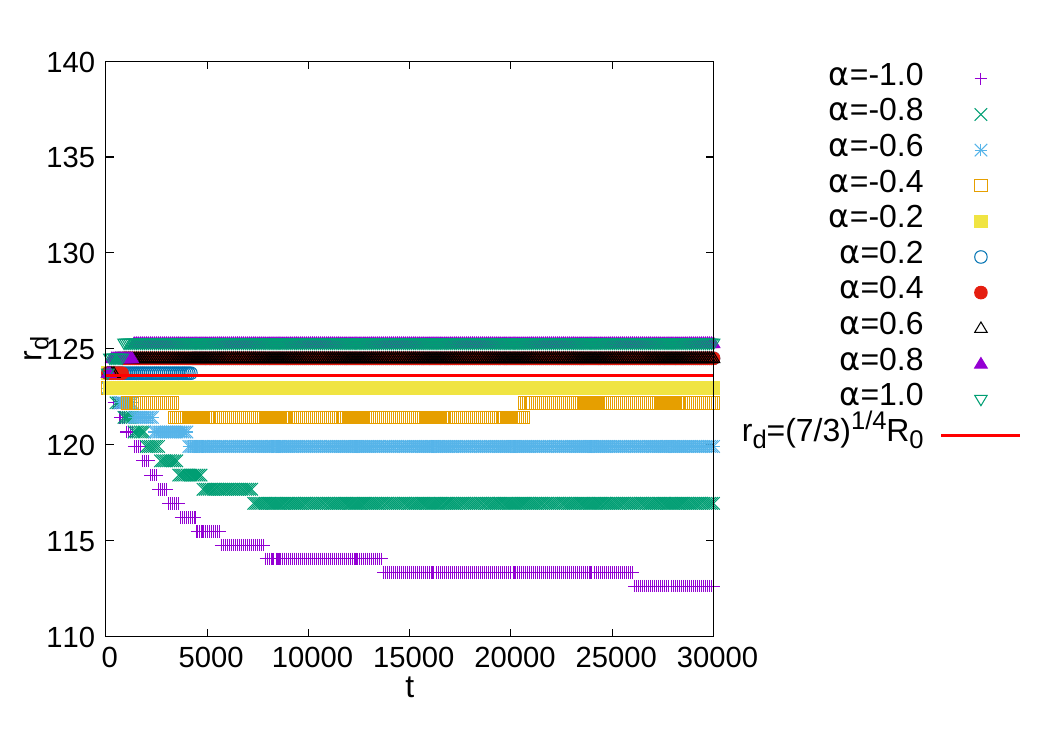}
		\subcaption{
			Homogeneous.
		}
		\label{fig_time_evolution_of_defect_position_homogeneous}
	\end{minipage}
	\caption{
		The time evolution of the $y$-component ($r_d$) of the position of the defect under
		(a) the homeotropic
		and
		(b) the homogeneous boundary conditions.
		Red horizontal line indicates the $y$-component of the equilibrium position
		in the case of passive nematic liquid crystal ($\alpha=0$).
	}
	\label{timestep_vs_r_d}
\end{figure*}

The time evolution of the distance $r_{d}$ between the origin and the defect core is shown in FIG.\ref{timestep_vs_r_d} for several $\alpha$'s
and both boundary conditions
(homeotropic
(FIG.\ref{fig_time_evolution_of_defect_position_homeotropic})
and
homogeneous
(FIG.\ref{fig_time_evolution_of_defect_position_homogeneous})).
It can be seen that the larger $|\alpha|$ is, the larger the deviation from the passive case is.

In FIG.
\ref{fig_velocity_on_y_axis_active_homeotropic_positive_alpha},
\ref{fig_velocity_on_y_axis_active_homeotropic_negative_alpha},
\ref{fig_velocity_on_y_axis_active_homogeneous_positive_alpha},
\ref{fig_velocity_on_y_axis_active_homogeneous_negative_alpha},
we can observe that the direction of the shift is identical to the direction of the flow field at the defect in each case.
This indicates that there is some sort of interaction between the defect and the flow, which drives the defect in the flow direction.

In the next section, we seek to explain these numerical results based on a theoretical framework.
 \subsection{Discussion based on an analytical expression of the velocity of the defect}
In the previous subsection (Sec.\ref{result_discussion_active_case}),
we have performed numerical simulations and found that
positions of $-1/2$ defects deviates from the ones in the case of a passive nematic liquid crystal.
To discuss the interaction between the flow field and the topological defect mentioned in Sec.\ref{result_discussion_active_case},
we use an analytical expression relating
the velocity of a defect with the profile of the nematic order parameter and the flow field at the core \cite{angheluta2021role, schimming2023kinematics, schimming2025analytical}.
As discussed in the following, this expression associates the velocity $\bm{v}=(v_x,v_y)^T$ of the defect with the force exerted on the defect.
The stationary position of the defect can be given by $\bm{v}=0$.
We apply this framework to the situation in our simulations,
and compare the results of simulations and theoretical calculations.
\subsubsection{The Halperin--Mazenko formalism applied to active nematic}
\label{the_halperin_mazenko_formalism}
The Halperin--Mazenko (HM) formalism
\cite{balian1981physics,liu1992defect,mazenko1997vortex,mazenko2001defect}
refers to a theoretical framework to analyze the motion of defect in the $O(n)$ model
focusing on the zeros of the order parameter.
Recently,
the HM formalism has been applied to two-dimensional nematic systems in \cite{angheluta2021role}.
The expression for three-dimensional nematic is derived in \cite{schimming2023kinematics}.
Schimming et al.
\cite{schimming2025analytical} have performed
a comparison between the theoretical calculation based on the HM formalism
and active nematic simulations in
circular confined domains, different from our setup.

Now let us consider a system including $N_d$ defects of charge $+1/2$ or $-1/2$, and focus on the $0$th defect
without loss of generality.
In the following discussion, we use complex variables defined as follows:
$z\equiv x+iy$ is the complex coordinate,
$\psi\equiv Q_{xx} + i Q_{xy}$ is the complex nematic order parameter.

The formula of the complex velocity $v^{(0)} = v_x^{(0)} + i v_y^{(0)}$ of the $0$th defect is given by
\cite{angheluta2021role}
\footnote{
	We use the formula in the complex form given in \cite{angheluta2021role} for the simplicity of the calculation.
	The one in terms of real variables can be found in \cite{schimming2023kinematics, schimming2025analytical}.
}
\begin{equation}
	\label{hm_formula}
	v^{(0)} = \left[
		\frac{
			-\partial_{\bar{z}}\bar{\psi}\partial_{t}\psi + \partial_{\bar{z}}\psi\partial_{t}\bar{\psi}
		}{
			\partial_{z}\psi\partial_{\bar{z}}\bar{\psi} - \partial_{z}\bar{\psi}\partial_{\bar{z}}\psi
		}
		\right]_{z=z^{(0)}},
\end{equation}
where the horizontal bar, e.g. $\bar{\psi}$, represents the complex conjugate,
and $[\cdots]_{z=z^{(0)}}$ indicates that ``$\cdots$'' is evaluated at the core of the $0$th defect.
In what follows, we indicate the quantities related to the $j$th defect by the superscript $\left(j\right)$.

As in \cite{angheluta2021role},
we assume the linear profile of $S$ near the core of the $0$th defect, namely, $S\propto|z-z^{(0)}|$ for $|z-z^{(0)}|\ll 1$.
By this assumption,
we can obtain the following expression for the velocity of the defect of charge $q^{(0)}=-1/2$,
see Appendix \ref{review_on_Halperin_Mazenko} for details:
\begin{equation}
	\label{v_defect_minus}
	v^{(0)} = \left[
		+8i\partial_{\bar{z}}\tilde{\Theta} + u - \lambda e^{2i\tilde{\Theta}}\partial_{z}\bar{u}
		\right]_{z=z^{(0)}},
\end{equation}
where $u\equiv u_x + i u_y$ is the complex velocity field.
$\tilde{\Theta}$ is the tilt of the director field
from which the contribution by the $0$th defect is removed:
\begin{equation}
	\label{Theta_tilde}
	\tilde{\Theta} = \frac{i}{2}\sum_{j=1}^{N_d - 1} q^{(j)}
	\log\left(\frac{\bar{z}-\overline{z^{(j)}}}{z - z^{(j)}}\right) + \Theta_{0},
\end{equation}
where $\Theta_{0}$ is a fixed external phase.
When the total topological charge is zero, which is the case in our simulations,
$\Theta_{0}$ is equal to the far field orientation.
We note that Eqs.(\ref{v_defect_minus}) and (\ref{Theta_tilde}) have already been derived in \cite{angheluta2021role}.

In the following calculations, $N_{d}=5$ and the defects are labeled as described in Fig.\ref{configuration_of_image_charges}.

\subsubsection{Analytical expression for the flow field around a $-1/2$ defect}
\label{analytical_expression_for_the_flow_field}
To evaluate Eq.(\ref{v_defect_minus}), we need to obtain the analytical expression for the flow field $u$.
The exact flow field will be obtained by solving the Stokes equation (Eq.(\ref{StokesEq_nondim})) under the given profile of $\psi$ (or $\bm{Q}$)
and nonslip boundary condition on the surface of the obstacle and the outer boundary.

However, as will be seen in Appendix \ref{review_on_flow_field} or \cite{ronning2022flow},
even the derivation of the flow field generated by an isolated defect is rather complicated, and then
it does not seem that exploring the flow field in our setup is feasible.
Thus, here let us make several assumptions to simplify the calculation:
\begin{enumerate}[label=(\roman*)]
	\item The flow is given simply by the superposition of the flows
		which each defect generates when they are isolated from the other defects.
		Consequently, the nonslip boundary condition is not imposed.
	\item In the force balance equation
		($\left(\text{r.h.s. of Eq.(\ref{v_defect_minus})}\right)=0$)
		for the $0$th defect, the flows generated by the $0$th and $1$st defects are taken into account and the flows by the $2$nd, $3$rd and $4$th defects are ignored,
		namely, the flows only by itself and nearest imaginary defect are taken into account.
\end{enumerate}

Strictly,
unlike the case of the orientation field $\Theta$, the flow field cannot be evaluated by the superposition of the flow generated by each defect
because the Stokes equation is not linear with respect to $\Theta$.
However, the assumption (i) has been adopted in \cite{schimming2025analytical},
which achieved a certain degree of success in explaining the results of numerical simulations.
This indicates that the correction arising from the non-linearity is not quite important in the calculation.

Next, let us comment on the assumption (ii), namely,
for the simplification,
we drop the flow induced by the defects other than the focused defect and its nearest image defect.
This can be justified by the following argument:
As discussed in the remaining of this subsection, the flow field generated by a defect decays as $\sim 1/r$ ($r$: distance from the defect core),
and the flow induced by the other defects has less contribution to the force balance equation than that from the nearest one.

Furthermore, because of its symmetric shape, the flow field generated by a $-1/2$ defect is zero at its core.
From this fact and the assumption (i), (ii),
we consider only the flow induced by the nearest image defect.

From the above discussions,
it is sufficient to obtain the flow field induced by an isolated $-1/2$ defect embedded in an otherwise uniform nematic field.
We need to obtain the flow field around the $j=1$ defect in the cases of homeotropic and homogeneous boundary condition.
The flow field around a defect is obtained in \cite{ronning2022flow} and we can use the result with minor modification related to the compressibility.
We present the procedure to solve Eq.(\ref{StokesEq_nondim}) in Appendix \ref{review_on_flow_field} and give only the final result here.
To simplify the calculation, we adopt the coordinate system where the $j=1$ defect is at the origin.
In that coordinate system, the complex velocity field generated by the $j=1$ defect is,
in terms of the polar coordinate $\left(r,\theta\right)$, given by
\begin{equation}
	\label{u_by_nearest_image_defect}
	u(r,\theta)
	=
e^{2i\Phi_{0}}
	\sum_{n=0}^{\infty}
	F_{n}\left(\alpha,\eta\right)
	\frac{
		e^{-2i\theta}
	}{
		r^{2n+1}
	},
\end{equation}
where
\begin{equation}
	\label{F_n_alpha_eta_1}
	F_{n}\left(\alpha,\eta\right)=
	S_{0}\alpha
	\left\{
		\frac{
			\left(2n-1\right)!!
		}{
			\left(2n\right)!!
		}
		\right\}^{2}
	\frac{2n+1}{2n-1}
	\left(4\eta\right)^{n}\left(n!\right)^{2}.
\end{equation}
In Eq.(\ref{u_by_nearest_image_defect}), $\Phi_{0}$ is the orientation of the director on the $x$-axis when the $j=1$ defect is located at the origin.
$\Phi_{0}=-\pi/4$ and $+\pi/4$ correspond to the homeotropic and homogeneous boundary conditions, respectively.
\subsubsection{Derivation of the force balance equation including the force exerted by the flow}
Now what we have to do is to evaluate the right hand side of Eq.(\ref{v_defect_minus})
and identify the stationary configuration of the defects by the condition $v^{(0)}=0$.
As already mentioned in Sec.\ref{analytical_expression_for_the_flow_field},
for the evaluation of the r.h.s. of Eq.(\ref{v_defect_minus}),
we adopt the coordinate whose origin is at the core of the $j=1$ defect.
In this coordinate system, the parameters in Eq.(\ref{Theta_tilde}) are given by
$N_{d}=5$,
$z^{(0)}=i\left(r_{d} - R_{0}^{2}/r_{d}\right)$,
$z^{(1)}=0$,
$z^{(2)}=-iR_{0}^{2}/r_{d}$,
$z^{(3)}=-i2R_{0}^{2}/r_{d}$,
$z^{(4)}=-i\left(r_{d} + R_{0}^{2}/r_{d}\right)$,
$q^{(0)}=q^{(1)}=q^{(3)}=q^{(4)}=-1/2$, $q^{(2)}=+2$, and
as shown in FIG.\ref{setup_fig},
$\Theta_{0}=0$ and $\Theta_{0}=\pi/2$ for cases of homeotropic and homogeneous boundary condition, respectively.

Each term in Eq.(\ref{v_defect_minus}) can be evaluated as follows:
We can calculate $\left.\partial_{\bar{z}}\tilde{\Theta}\right|_{z=z^{(0)}}$ using Eq.(\ref{Theta_tilde}) to obtain
\begin{equation}
	\label{partial_zbar_Theta_tilde}
	\left.\partial_{\bar{z}}\tilde{\Theta}\right|_{z=z^{(0)}}
	=\frac{i}{2}\sum_{j=1}^{N_{d}-1}\frac{
		q^{(j)}
	}{
		\overline{z^{(0)}}-\overline{z^{(j)}}
	}
	=-\frac{1}{2}\sum_{j=1}^{N_{d}-1}\frac{q^{(j)}}{r_{0j}},
\end{equation}
where $r_{jk}\equiv\left|z^{(j)}-z^{(k)}\right|$.
Eq.(\ref{partial_zbar_Theta_tilde}) does not depend on $\Theta_{0}$ and applies to both homeotropic and homogeneous cases.
$\left.e^{2i\tilde{\Theta}}\right|_{z=z^{(0)}}$ can be calculated using Eq.(\ref{Theta_tilde})
and the result is $\left.e^{2i\tilde{\Theta}}\right|_{z=z^{(0)}}=ie^{2i\Theta_{0}}$, namely,
\begin{equation}
	\label{exp_factor}
	\left.e^{2i\tilde{\Theta}}\right|_{z=z^{(0)}}=
	\begin{cases}
		+i & (\text{homeotropic})\\
		-i & (\text{homogeneous})
	\end{cases}.
\end{equation}
$\partial_{z}\overline{u}$ can be calculated by rewriting $\partial_{z}$ in terms of $\left(r,\theta\right)$,
namely, $\partial_{z}=\left(e^{-i\theta}/2\right)\left\{\partial_{r}-\left(i/r\right)\partial_{\theta}\right\}$.
The result is
\begin{equation}
	\label{partial_z_ubar}
	\left.\partial_{z}\overline{u}\right|_{z=z^{(0)}}=
-ie^{2i\Phi_{0}}
	\sum_{n=0}^{\infty}F_{n}\left(\alpha,\eta\right)
	\left(-n+\frac{1}{2}\right)\frac{1}{r_{01}^{2n+2}}.
\end{equation}
Substituting Eqs.(\ref{partial_zbar_Theta_tilde}), (\ref{exp_factor}) and (\ref{partial_z_ubar}) into Eq.(\ref{v_defect_minus}),
we obtain the force balance equation for the $j=0$ defect:
\begin{equation}
	\label{force_balance_equation}
	\begin{split}
		&-4\sum_{j=1}^{N_{d}-1}\frac{
			q^{(j)}
		}{
			r_{0j}
		}\\
		&+ \sum_{n=0}^{\infty}F_{n}\left(\alpha,\eta\right)
		\left\{
			\pm\frac{1}{r_{01}^{2n+1}}
			-\lambda
			\left(
			n-\frac{1}{2}
			\right)
			\frac{1}{r_{01}^{2n+2}}
			\right\}
		=0,
	\end{split}
\end{equation}
where $+$ and $-$ correspond to homeotropic and homogeneous boundary condition, respectively.
When $\alpha=0$, Eq.(\ref{force_balance_equation}) reduces to the force balance equation ($\sum_{j=1}^{N_{d}-1}q^{(j)}/r_{01}=0$) in the passive case.
In the next subsection (Sec.\ref{symmetry_of_equation_under_rotation_of_boundary_condition}),
Eq.(\ref{force_balance_equation}) is solved numerically, and the results are compared with the simulations.

\begin{figure*}[btp]
	\begin{minipage}[b]{0.49\linewidth}
		\centering
		\includegraphics[width=0.99\linewidth]{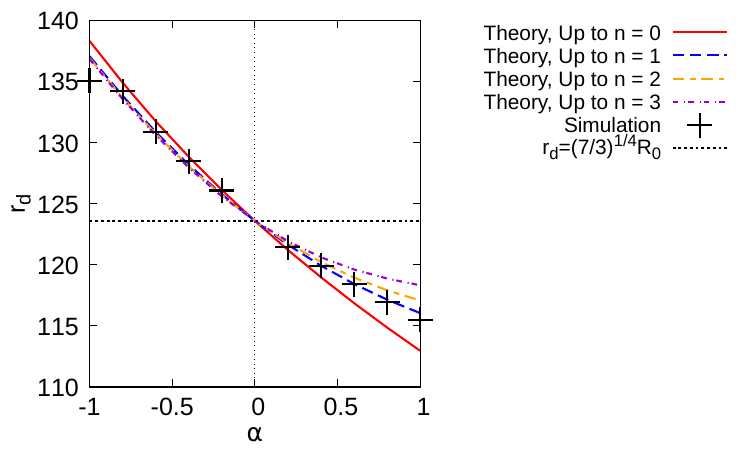}
		\subcaption{
			Homeotropic.
		}
		\label{result_compare_simulation_and_theory_homeotropic}
	\end{minipage}
	\begin{minipage}[b]{0.49\linewidth}
		\centering
		\includegraphics[width=0.99\linewidth]{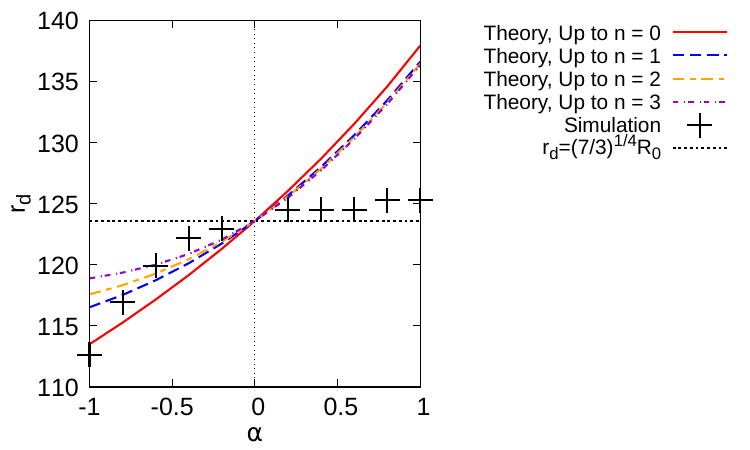}
		\subcaption{
			Homogeneous.
		}
		\label{result_compare_simulation_and_theory_homogeneous}
	\end{minipage}
	\caption{
		Dependence of the stationary position $r_{d}$ on the activity $\alpha$.
		Lines are the numerical solutions of Eq.(\ref{force_balance_equation}),
		black cross points represent the results of the numerical simulations, and
		the horizontal dashed line indicates the defect position in the passive case.
		In the computation of Eq.(\ref{force_balance_equation}),
		we truncate the infinite summation $\sum_{n=0}^{\infty}$ at $n=0$, $1$, $2$ and $3$.
	}
	\label{result_compare_simulation_and_theory}
\end{figure*}
 \subsubsection{Comparison between simulation results and theoretical calculations}
\label{symmetry_of_equation_under_rotation_of_boundary_condition}
The force balance equation of the defect (Eq.(\ref{force_balance_equation}))
can be solved by the bisection method, and we can obtain the dependence of $r_{d}$ on $\alpha$;
see the curves in FIG.\ref{result_compare_simulation_and_theory}.
In the computation of Eq.(\ref{force_balance_equation}),
we truncate the summation of infinite series at $n=0,1,2,3$
and plot four theoretical curves in FIG.\ref{result_compare_simulation_and_theory}.

To compare the theoretical calculations with the simulation results, we also plot the stationary values of $r_{d}$ for each $\alpha$;
see the red cross points in FIG.\ref{result_compare_simulation_and_theory}.
These are the final stationary values of $r_{d}$ in FIG.\ref{timestep_vs_r_d}.

In FIG.\ref{result_compare_simulation_and_theory}, we can observe that
the theoretical calculations agree well with the simulation results
in the homeotropic case (FIG.\ref{result_compare_simulation_and_theory_homeotropic}),
which indicates the validity of our theoretical calculations.
In the homogeneous case (FIG.\ref{result_compare_simulation_and_theory_homogeneous}), however, the agreement is
not good, especially in the $\alpha>0$ regime, which
indicates that our theoretical framework does not work well there.

It is not clear why our calculations are not valid in the homogeneous case.
However, we can explain why the theoretical curve in the homeotropic and the homogeneous cases
have almost symmetric shape with respect to $\alpha=0$, but the simulation results do not.

To have an insight into the difference arising from the boundary condition,
let us discuss the symmetry of the basic equations under the rotation of $\bm{n}$ by $\pi/2$
, which renders the homeotropic (homogeneous) alignment at the boundary to homogeneous (homeotropic).

When $\bm{n}$ is rotated by $\pi/2$, $\bm{Q}$ changes to $-\bm{Q}$.
In the passive case, all terms in the basic equation (Eq.(\ref{model_A})) are odd with respect to $\bm{Q}$.
The basic equation is thus invariant under the rotation of $\bm{n}$, and
the equilibrium defect positions are identical in the two cases (see FIG.\ref{result_passive}).

In the active case, the theoretical curves for different boundary conditions in Fig.\ref{result_compare_simulation_and_theory}
appear symmetric with respect to $\alpha=0$ (although they are not as we will discuss below), and let us consider as well the effect of the sign of $\alpha$.
The Stokes equation (Eq.(\ref{StokesEq_nondim})), and thus its solution $\bm{u}$ are invariant
under the simultaneous transformation of $\bm{Q}$ and $\alpha$ to $-\bm{Q}$ and $-\alpha$.
Hence under this transformation the $\lambda\bm{E}$ term in eq.(\ref{EdwardsBerisEq_nondim}) is invariant, although all the other terms change sign.
Therefore, in contrast to the above-mentioned passive case, the $\pi/2$-rotation of the stationary profile of $\bm{n}$
for the homeotropic boundary condition and a given $\alpha$ does not yield a stationary solution of eq.(\ref{EdwardsBerisEq_nondim}) for
the homogeneous boundary condition and $-\alpha$. The different parity of the $\lambda\bm{E}$ term
under this transformation is responsible for the asymmetry of the simulation results with respect to $\alpha=0$
observed in Fig.\ref{result_compare_simulation_and_theory}.

In addition, the asymmetry attributable to the $\lambda$-term
can be seen in the force balance equation (Eq.(\ref{force_balance_equation})).
Namely, while the first term $\pm 1/r_{01}^{2n+1}$ changes its sign
as the boundary condition changes, the second term $-\lambda\left(n-1/2\right)/r_{01}^{2n+2}$ does not.
However, the contributions from the $\lambda$-term are higher-order of $1/r_{01}$ in the theoretical calculation
(see Eq.(\ref{force_balance_equation})).
As a result, the theoretical curves are almost symmetric in the two cases
(see FIG.\ref{result_compare_simulation_and_theory}) and
fail to reproduce the simulation results, which are quite asymmetric in the two cases.

Furthermore, let us discuss the point
that the agreement between theory and simulation is not improved when the higher-order terms are included;
see $\alpha > 0$ regime of FIG.\ref{result_compare_simulation_and_theory_homeotropic}, where
the theory and simulation show best agreement when the infinite summation in Eq.(\ref{force_balance_equation}) is truncated up to $n=1$.
Although one might expect that the theoretical curve gets closer to the simulation results
when the higher-order ($n=2,3,\cdots$) terms are included, it is not the case.
Our theoretical framework is based on several approximations listed in Sec.\ref{analytical_expression_for_the_flow_field}.
If the difference between the theory and simulation arising from those approximations decreased by including the higher-order terms, the theoretical curve
would get closer to the simulation results.
However, it is not the case in our framework, and we cannot expect that the agreement increases.
Our result of homeotropic case (FIG.\ref{result_compare_simulation_and_theory_homeotropic}) indicates
that the approximations do not improve when higher-order terms are included.
We would like to leave further elaboration of our theory for future studies.

As already mentioned above, it is not clear why the theoretical curve and the simulation results
agree well in the homeotropic case and do not in the homogeneous case,
especially in the contractile ($\alpha>0$) regime.
Understanding why the deviation is suppressed in the contractile ($\alpha>0$) regime
under the homogeneous boundary condition is an open question and left to future work.
  \section{Conclusion \& Outlook}
\label{conclusion}
In this paper, we have investigated the stationary configuration of topological defects in an active nematic around a circular obstacle.

First, we have performed numerical simulations based on a continuum model.
As a result, we found that the defect configuration deviates from the one observed in passive nematic liquid crystals.
This deviation is attributed to the flow induced by the active stress.
The magnitude of the deviation can be controlled by the strength $|\alpha|$ of active stress, namely,
the larger the strength $|\alpha|$ of activity is, the larger deviation from the passive configuration is observed.

Furthermore, we have performed a calculation based on an analytically tractable theory
to explain the simulation results.
For that purpose, we employed the analytical expression relating the defect velocity $v$
with the nematic order parameter and the flow field, which had been derived based on the Halperin-Mazenko formalism in previous studies.
It was applied to the situation of our simulations and the stationary configuration was identified by the condition $v=0$.
Our theoretical calculation has
qualitatively
reproduced our simulation results, which supports the validity of the Halperin--Mazenko approach
in the analysis of active nematic systems.

Finally, let us mention several potential directions for future studies.
The first one concerns the compressibility.
In this paper, we have employed a simplified framework for the compressibility, namely, we have assumed the constant density and pressure
by relaxing the mass conservation condition.
If the mass conservation is imposed, one has to solve it to obtain the density field and
then, obtain the pressure via a type of equation of state.
However, the form of the equation of state does not seem to be trivial.
Exploring the equation of state and performing the numerical simulation considering the spatial variation of the density
will be a intriguing topic of future work.

Next one is about the turbulent behavior of active nematic.
In this paper, we have investigated the small-$|\alpha|$ regime, because
we have focused on the stationary defect configuration and its theoretical analysis.
For the large-$|\alpha|$ regime, the behavior becomes turbulent.
Predicting the onset of turbulent behavior by a stability analysis and investigating the active-nematic turbulence around a circular obstacle
will be a direction of future study.

 \section{Acknowledgement}
\label{acknowledgement}
We thank Mr. Ryo Ienaga, Professor Kazusa Beppu and Professor Yusuke T. Maeda for productive discussions on their
experimental studies that inspired this study.
We are also grateful to Professor Holger Stark for fruitful discussions.
H.M. was supported by JST SPRING, Grant Number JPMJSP2136.
J.F. was supported in part by JSPS KAKENHI (Grants No. JP21H01049 and No. JP23K20831).
This work was supported by the JSPS Core-to-Core Program ``Advanced core-to-core network for the physics of self-organizing active matter (JPJSCCA20230002)''
 \begin{appendices}
\section{Outline of the derivation of basic equations in $\left(\xi,\theta\right)$-coordinate}
\label{equations_in_xi_theta_coordinate}
As mentioned in Sec.\ref{model}, we employ the two dimensional polar coordinate and choose $Q_{rr}$ and $Q_{r\theta}$ as two independent components of $\bm{Q}$.
Thus, equations we have to solve are the $rr$- and $r\theta$-components of eq.(\ref{EdwardsBerisEq_nondim}) and $r$- and $\theta$-components of eq.(\ref{StokesEq_nondim}).
It is a straightforward task to write them in terms of
$Q_{rr}$, $Q_{r\theta}$, $u_{r}$ and $u_{\theta}$ with spatial coordinate $r$ and $\theta$.
As already mentioned, we perform one more coordinate transformation from $r$ to $\xi$.
Using the relations $r=R_{0}e^{\xi}$, $\partial_{r}=(e^{\xi}R_{0})^{-1}\partial_{\xi}$,
$\bm{\nabla}^2=(e^{\xi}R_{0})^{-2}(\partial_{\xi}\partial_{\xi}+\partial_{\theta}\partial_{\theta})$, etc...,
we can eliminate $r$ from the equations and obtain the basic equation of our simulation in $(\xi,\theta)$-coordinate:

\begin{widetext}
	\begin{align}
		\partial_{{t}}Q_{rr} =&
		-\frac{e^{-\xi}}{{R}_0}{u}_{r}Q_{rr,\xi}
		-\frac{e^{-\xi}}{{R}_0}{u}_{\theta}\left(Q_{rr,\theta}-2Q_{r\theta}\right)
		+\frac{e^{-\xi}}{{R}_0}\frac{\lambda}{2}\left\{{u}_{r,\xi}-({u}_{r}+{u}_{\theta,\theta})\right\} \notag \\
		& + \frac{e^{-\xi}}{{R}_0}\left(-{u}_{\theta,\xi}+{u}_{r,\theta}-{u}_{\theta}\right)Q_{r\theta}
		+\frac{e^{-2\xi}}{{R}_0^2}(\partial_{\xi\xi}+\partial_{\theta\theta}) Q_{rr}
		-\frac{4e^{-2\xi}}{{R}_0^2}(Q_{r\theta,\theta}+Q_{rr})
		+ \left(1-{\rm{tr}}{\bm Q}^2\right)Q_{rr}
		\label{EdwardsBerisEq_rr3}\\
		\partial_{{t}}Q_{r\theta} =&
		-\frac{e^{-\xi}}{{R}_0}{u}_{r}Q_{r\theta,\xi}
		-\frac{e^{-\xi}}{{R}_0}{u}_{\theta}\left(Q_{r\theta,\theta}+2Q_{rr}\right)
		+\frac{e^{-\xi}}{{R}_0}\frac{\lambda}{2}\left\{ {u}_{\theta,\xi}+({u}_{r,\theta}-{u}_{\theta})\right\} \notag \\
		& +\frac{e^{-\xi}}{{R}_0} \left({u}_{\theta,\xi}-{u}_{r,\theta}+{u}_{\theta}\right)Q_{rr}
		+\frac{e^{-2\xi}}{{R}_0^2}(\partial_{\xi\xi}+\partial_{\theta\theta}) Q_{r\theta}
		+\frac{4e^{-2\xi}}{{R}_0^2}(Q_{rr,\theta}-Q_{r\theta})
		+ \left(1-{\rm{tr}}{\bm Q}^2\right)Q_{r\theta}
		\label{EdwardsBerisEq_rtheta3}\\
		0=&-{u}_{r}
		+\frac{e^{-2\xi}}{{R}_0^2}{\eta}\left\{
			(\partial_{\xi\xi}+\partial_{\theta\theta}){u}_{r}-{u}_{r}-2{u}_{\theta,\theta}
			\right\}
		+\frac{e^{-\xi}}{{R}_0}{\alpha}\left(
		Q_{rr,\xi}+Q_{r\theta,\theta}+2Q_{rr}
		\right)
		\label{StokesEq_r3}\\
		0=&-{u}_{\theta}
		+\frac{e^{-2\xi}}{{R}_0^2}\eta\left\{
			(\partial_{\xi\xi}+\partial_{\theta\theta}){u}_{\theta}-{u}_{\theta}+2{u}_{r,\theta}
			\right\}
		+\frac{e^{-\xi}}{{R}_0}{\alpha}\left(
		Q_{r\theta,\xi}-Q_{rr,\theta}+2Q_{r\theta}
		\right)
		\label{StokesEq_theta3}
	\end{align}
\end{widetext}
where $f_{,\xi}$, $f_{,\theta}$, $\cdots$ denote the partial derivatives $\partial_\xi f$, $\partial_\theta f$, $\cdots$.
 	\section{Outline of the procedure to solve the Stokes equation}
\label{how_to_solve_Stokes_eq}
We present the outline of the procedure to solve the Stokes equation (Eq.(\ref{StokesEq_nondim})) numerically.

\begin{figure}[btp]
	\centering
	\includegraphics[width=0.99\linewidth]{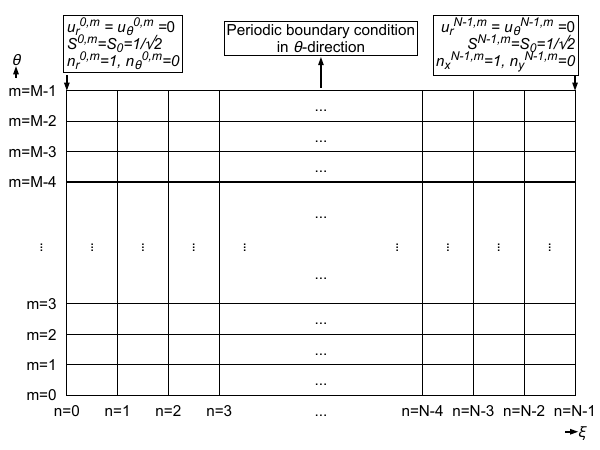}
	\caption{
		The schematic of $(\xi,\theta)$-coordinate system.
	}
	\label{xi_theta_coordinate}
\end{figure}

The Stokes equation in $(\xi,\theta)$-coordinate is shown in Appendix \ref{equations_in_xi_theta_coordinate} (Eq.(\ref{StokesEq_r3}) and
(\ref{StokesEq_theta3})).
Let us denote the total lattice numbers in each ($\xi$ and $\theta$) direction by $N$ and $M$, respectively.
By applying the finite differential scheme on Eqs.(\ref{StokesEq_r3}) and (\ref{StokesEq_theta3}),
we obtain $2(N-2)M$ linear equations
\footnote{
	The factor 2 arises from two ($r$ and $\theta$) components.
	Since the velocity at the inner and outer boundaries is fixed by the nonslip and impermeability conditions,
	the number of Stokes equations is $(N-2)M$.
} whose unknowns are $u_r$ and $u_{\theta}$ defined at each lattice point of $(\xi,\theta)$-space (see FIG.\ref{xi_theta_coordinate}).
These equations can be rewritten as $A\bm{x}=\bm{b}$,
where $A$ is the coefficient matrix, $\bm{x}$ is the unknown vector whose entries are $u_r$
and $u_{\theta}$ at each lattice point, and $\bm{b}$ is the constant vector that corresponds to the active stress of the Stokes equation.
One can identify the entries of $A$ and $\bm{b}$ by straightforward calculations, and solve the equation $A\bm{x}=\bm{b}$
by a solver for linear algebra problems (e.g. C++ Eigen \cite{cpp_eigen}, which we employed.).
 \section{Review on the calculation of the Halperin-Mazenko formula under the linear core approximation}
\label{review_on_Halperin_Mazenko}
We review the procedure \cite{angheluta2021role} for deriving Eq.(\ref{v_defect_minus}) from Eq.(\ref{hm_formula}).
Although both cases of $q^{(0)}=+1/2$ and $q^{(0)}=-1/2$ are discussed in \cite{angheluta2021role},
let us restrict our discussion to $q^{(0)}=-1/2$, which is required for our discussions.

By assuming the linear profile of the scalar order parameter near the $0$-th defect core,
i.e. $S=C|z-z^{(0)}|$ ($C$: a real constant) for $|z-z^{(0)}|\ll 1$,
we can write the complex order parameter as
\begin{equation}
	\psi = C\left|z-z^{(0)}\right|e^{2i\Theta}
	\quad\text{for}\quad
	\left|z-z^{(0)}\right|\ll 1,
\end{equation}
where $\Theta$ is given by
\begin{equation}
	\Theta=\frac{i}{2}\sum_{j=0}^{N_d-1}
	q^{(j)}\log\left[
	\frac{\bar{z}-\overline{z^{(j)}}}{z-z^{(j)}}
	\right]
	+\Theta_0.
\end{equation}
$q^{(j)}$ is the charge of the $j$-th defect and $\Theta_{0}$ is the external fixed phase.
Furthermore, $\psi$ can be rewritten as follows:
\begin{equation}
	\label{psi}
	\psi = C\left\{ \bar{z} - \overline{z^{(0)}}\right\} e^{2i\widetilde{\Theta}},
\end{equation}
where $\widetilde{\Theta}$ is defined by
\begin{equation}
	\widetilde{\Theta}
	=\frac{i}{2}\sum_{j=1}^{N_d -1}
	q^{(j)}\log\left[
	\frac{\bar{z}-\overline{z^{(j)}}}{z-z^{(j)}}
	\right]
	+\Theta_0.
\end{equation}
Using Eq.(\ref{psi}), we obtain
$\left.\partial_{z}\psi\right|_{z=z^{(0)}}=0$,
$\left.\partial_{\bar{z}}\overline{\psi}\right|_{z=z^{(0)}}=0$,
$\left.\partial_{\bar{z}}\psi\right|_{z=z^{(0)}}=\left.Ce^{2i\widetilde{\Theta}}\right|_{z=z^{(0)}}$,
$\left.\partial_{z}\overline{\psi}\right|_{z=z^{(0)}}=\left.Ce^{-2i\widetilde{\Theta}}\right|_{z=z^{(0)}}$
and
$\left.\partial_{z}\partial_{\bar{z}}\psi\right|_{z=z^{(0)}}
=i2C\left[\left(\partial_{z}\widetilde{\Theta}\right)e^{2i\widetilde{\Theta}}\right]_{z=z^{(0)}}$.

$\left.\partial_{t}\psi\right|_{z=z^{(0)}}$ and $\left.\partial_{t}\overline{\psi}\right|_{z=z^{(0)}}$
can be evaluated by the Edwards-Beris equation (Eq.(\ref{EdwardsBerisEq_nondim})).
Rewriting Eq.(\ref{EdwardsBerisEq_nondim}) in terms of complex variables and setting $z=z^{(0)}$, we obtain
$\left.\partial_{t}\psi\right|_{z=z^{(0)}}
= \left[
	-u\partial_{z}\psi
	-\bar{u}\partial_{\bar{z}}\psi
	+\lambda\partial_{\bar{z}}u
	+4\partial_{z}\partial_{\bar{z}}\psi
	\right]_{z=z^{(0)}}$.
$\left.\partial_{t}\overline{\psi}\right|_{z=z^{(0)}}$ can be
obtained by taking the complex conjugate of $\left.\partial_{t}\psi\right|_{z=z^{(0)}}$.

Substituting the above results into Eq.(\ref{hm_formula}), we obtain
\begin{equation}
	v^{(0)} =
	\left[ +8i\partial_{\bar{z}}\widetilde{\Theta} + u - \lambda e^{2i\widetilde{\Theta}}\partial_{z}\bar{u}\right]_{z=z^{(0)}},
\end{equation}
where we have assumed that $C=1$ to obtain the same expression as in \cite{angheluta2021role}.
 	\section{Review on the derivation of the flow field around a topological defect}
\label{review_on_flow_field}
Solving the Stokes equation (Eq.(\ref{StokesEq_nondim})) under the existence of a topological defect requires a lengthy calculation,
but we can utilize the result of a previous study \cite{ronning2022flow} with a minor modification.

The Stokes equation (Eq.(\ref{StokesEq_nondim})) can be rewritten as
\begin{equation}
	\label{stokes_eq_modified}
	\left\{
		\bm{\nabla}^2 - \left(\frac{1}{\sqrt{\eta}}\right)^{2}
		\right\}
	\bm{u} = -\frac{\alpha}{\eta}\bm{\nabla}\cdot\bm{Q}.
\end{equation}
Eq.(\ref{stokes_eq_modified}) can be solved formally by introducing the Green function
that satisfies $\{\bm{\nabla}^2 - (1/\sqrt{\eta})^2\}G(\bm{r},\bm{r}^{\prime})=-\delta(\bm{r}-\bm{r}^{\prime})$
under a boundary condition $G(\bm{r},\bm{r}^{\prime})=0$ as $|\bm{r}|\rightarrow\infty$.
It is explicitly given as
$G(\bm{r},\bm{r}^{\prime})=(2\pi)^{-1}K_{0}\left(\left|\bm{r}-\bm{r}^{\prime}\right|/\sqrt{\eta}\right)$
\cite{arfken2011mathematical},
where $K_{0}$ is the modified Bessel function of the second kind.
Using this Green function, the solution to Eq.(\ref{stokes_eq_modified}) can be written as
\begin{equation}
	\label{u_Green_func}
	\begin{split}
		\bm{u}(\bm{r})
= \frac{\alpha}{2\pi\eta}\iint
		d\bm{r}^{\prime}K_{0}\left(
		\frac{|\bm{r}-\bm{r}^{\prime}|}{\sqrt{\eta}}
		\right)
		\bm{\nabla}^{\prime}\cdot\bm{Q}(\bm{r}^{\prime})
	\end{split}
\end{equation}
where the integral is taken over the whole space.

\begin{figure}[btp]
	\begin{minipage}[b]{0.45\linewidth}
		\centering
		\includegraphics[width=0.99\linewidth]{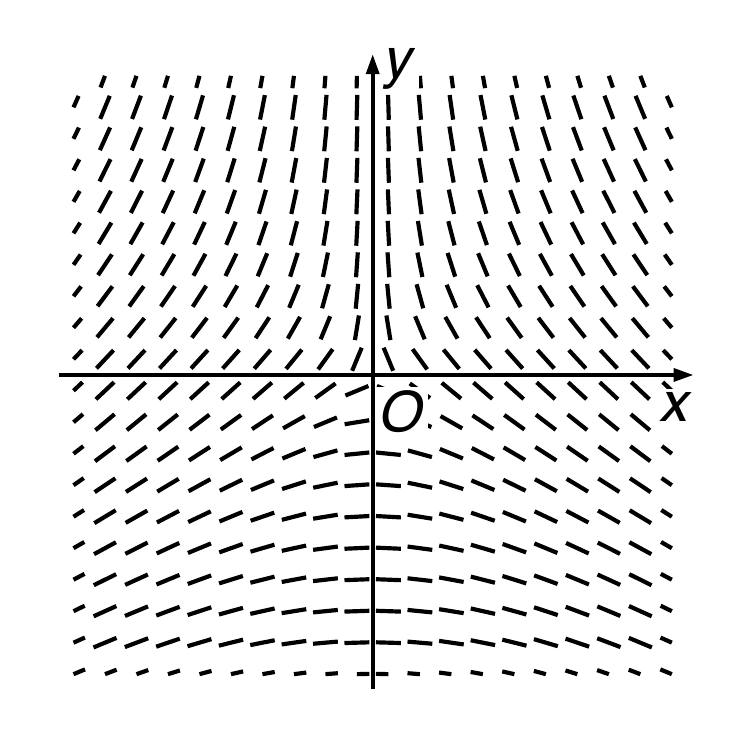}
		\subcaption{$\Phi_{0}=-\pi/4$}
		\label{minus_one_half_defect_upward}
	\end{minipage}
	\begin{minipage}[b]{0.45\linewidth}
		\centering
		\includegraphics[width=0.99\linewidth]{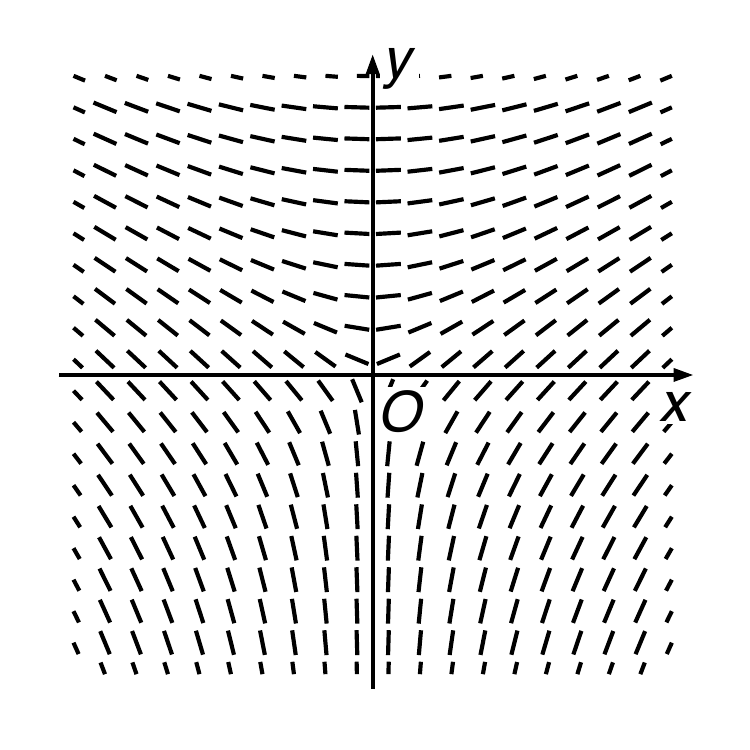}
		\subcaption{$\Phi_{0}=+\pi/4$}
		\label{minus_one_half_defect_downward}
	\end{minipage}
	\caption{
		The director field $\Theta = (-1/2)\theta + \Phi_{0}$
		for (a)$\Phi_{0}=-\pi/4$
		and (b)$\Phi_{0}=+\pi/4$.
		The black bar is the director given by $\bm{n} = \left(\cos\Theta,\sin\Theta\right)$.
	}
	\label{topological_defects}
\end{figure}

Let us now consider a $-1/2$ defect whose core is at the origin, namely,
we consider the orientational field given by $\Theta=(-1/2)\theta + \Phi_{0}$,
where $\Phi_{0}$ is the orientation of the director on the $x$-axis ($\theta = 0$).
We restrict our discussion to the cases of $\Phi_{0}=\pm\pi/4$, which is sufficient for the calculation in the main text.
The director fields around the origin in the case of $\Phi_{0}=-\pi/4$ and $+\pi/4$
are shown in FIG.\ref{minus_one_half_defect_upward} and \ref{minus_one_half_defect_downward}, respectively.

For $\Phi_{0}=\pm\pi/4$, $\bm{\nabla}\cdot\bm{Q}(\bm{r})$ is given by
\begin{equation}
	\label{divQ_around_defect}
	\begin{split}
		&\bm{\nabla}\cdot\bm{Q}(\bm{r})=\\
		&\begin{cases}
			0 & (\text{at the core})\\
			e^{2i\left(\Phi_{0}+\frac{\pi}{4}\right)}\dfrac{S_{0}}{r}\left\{
				\sin(2\theta)\bm{e}_x + \cos(2\theta)\bm{e}_y
				\right\}
			& (\text{away from the core}).
		\end{cases}
	\end{split}
\end{equation}
Using Eq.(\ref{divQ_around_defect}), Eq.(\ref{u_Green_func}) can be evaluated as follows:
\begin{widetext}
	\begin{equation}
		\label{u_Green_func_2}
		\begin{split}
			\bm{u}(\bm{r})
			=&e^{2i\left(\Phi_{0}+\frac{\pi}{4}\right)}\frac{\alpha}{2\pi\eta}
			\int_{r_c}^{\infty}dr^{\prime}\int_{0}^{2\pi}d\theta^{\prime}
			K_{0}\left(
			\frac{|\bm{r}-\bm{r}^{\prime}|}{\sqrt{\eta}}
			\right)
			S_{0}\left\{
				\sin(2\theta^{\prime})\bm{e}_x + \cos(2\theta^{\prime})\bm{e}_y
				\right\}\\
\simeq &e^{2i\left(\Phi_{0}+\frac{\pi}{4}\right)}\frac{\alpha}{2\pi\eta}
			\iint d\bm{r}^{\prime}
			K_{0}\left(
			\frac{|\bm{r}-\bm{r}^{\prime}|}{\sqrt{\eta}}
			\right)
			\frac{S_{0}}{r^{\prime}}\left\{
				\sin(2\theta^{\prime})\bm{e}_x + \cos(2\theta^{\prime})\bm{e}_y
				\right\},
		\end{split}
	\end{equation}
\end{widetext}
where $r_{c}$ is the radius of the defect core and
we have taken the limit $r_{c}\rightarrow 0$.

Finally,
let us change the variable of integral from $\bm{r}^{\prime}$
to $\bm{r}^{\prime\prime}\equiv\bm{r}^{\prime}-\bm{r}$, and
introduce the complex variables $u=u_x + i u_y$,
$z\equiv x+iy$, $z^{\prime}\equiv x^{\prime}+iy^{\prime}$,
$r^{\prime\prime}e^{i\theta^{\prime\prime}}=r^{\prime\prime}\hat{z}\equiv z^{\prime}-z$.
With these variables, Eq.(\ref{u_Green_func_2}) can be rewritten as
\begin{widetext}
	\begin{equation}
		\label{u_Green_func_3}
		\begin{split}
			u(z,\bar{z}) =
			e^{2i\left(\Phi_{0}+\frac{\pi}{4}\right)} i\frac{S_{0}\alpha}{2i\pi\eta}
			\int_{0}^{\infty} dr^{\prime\prime}r^{\prime\prime}
			\oint_{\gamma}\frac{d\hat{z}}{\hat{z}}K_{0}\left(
			\frac{r^{\prime\prime}}{\sqrt{\eta}}
			\right)
			\frac{1}{z+r^{\prime\prime}\hat{z}}
			\sqrt{
				\frac{
					\bar{z} + r^{\prime\prime}\hat{z}^{-1}
				}{
					z + r^{\prime\prime}\hat{z}
				}
			},
		\end{split}
	\end{equation}
\end{widetext}
where $\gamma$ is the unit circle in the complex plane whose center is at the origin.

The integral of $\hat{z}$ in Eq.(\ref{u_Green_func_3}) has been performed in Appendix C of \cite{ronning2022flow}, and
the far-field asymptotic flow is given by
\begin{widetext}
	\begin{equation}
		\label{u_far_field}
u(r,\theta)
=
			-e^{2i\left(\Phi_{0}+\frac{\pi}{4}\right)} i\frac{S_{0}\alpha}{\sqrt{\eta}}\sum_{n=0}^{\infty}\left\{
				\frac{(2n-1)!!}{(2n)!!}
				\right\}^{2}
			\frac{2n+1}{2n-1}
			e^{-2i\theta}
			\left(\frac{\sqrt{\eta}}{r}\right)^{2n+1}
			\int_{0}^{\infty}dxK_{0}\left(x\right)
			x^{2n+1},
\end{equation}
\end{widetext}
where we have changed the variable in the integral by $x=r^{\prime\prime}/\sqrt{\eta}$.
The integral in Eq.(\ref{u_far_field}) can be evaluated as
$\int_{0}^{\infty}dxK_{0}(x)x^{2n+1} = 2^{2n}\left\{\Gamma(n+1)\right\}^{2}=4^{n}(n!)^{2}$,
and Eq.(\ref{u_far_field}) can be rewritten as follows:
\begin{equation}
	\label{u_far_field_expand}
	u(r,\theta)
	=
	-e^{2i\left(\Phi_{0}+\frac{\pi}{4}\right)} i
	\sum_{n=0}^{\infty}
	F_{n}\left(\alpha,\eta\right)
	\frac{
		e^{-2i\theta}
	}{
		r^{2n+1}
	},
\end{equation}
where
\begin{equation}
	\label{F_n_alpha_eta_2}
	F_{n}\left(\alpha,\eta\right)=
	S_{0}\alpha
	\left\{
		\frac{
			\left(2n-1\right)!!
		}{
			\left(2n\right)!!
		}
		\right\}^{2}
	\frac{2n+1}{2n-1}
	\left(4\eta\right)^{n}\left(n!\right)^{2}.
\end{equation}

 \end{appendices}
\bibliographystyle{plainurl}
\bibliographystyle{plainnat}
\bibliography{../../References/ActiveNematics/reference}
\bibliographystyle{unsrt}
\end{document}